%% file: mainExperiment.tex
\documentclass[journal]{IEEEtran}

\usepackage{lipsum}
\usepackage{amsmath,amssymb,amsfonts,mathrsfs, mathtools, dsfont, mathrsfs, amsthm, bm, bbm}
\usepackage[dvipsnames]{xcolor}
\DeclareMathAlphabet{\mathpzc}{OT1}{pzc}{m}{it}
\DeclareSymbolFontAlphabet{\amsmathbb}{AMSb}%
\usepackage{csvsimple,booktabs}
\usepackage{float, enumitem}
\usepackage{mdframed}
\usepackage{graphicx, epsfig}
\usepackage{balance}
\usepackage{multirow}
\usepackage[caption=false]{subfig}
\usepackage{cite}
\usepackage[T1]{fontenc}
\usepackage[normalem]{ulem}

\usepackage{tkz-euclide,tikzscale}
\usetikzlibrary{arrows,circuits.ee.IEC}
\tikzset{ac source/.style={
  circuit symbol lines,
  circuit symbol size = width 2 height 2,
  shape = generic circle IEC,
  /pgf/generic circle IEC/before background={
    \pgfpathmoveto{\pgfpoint{-0.8pt}{0pt}}
    \pgfpathsine{\pgfpoint{0.4pt}{0.4pt}}
    \pgfpathcosine{\pgfpoint{0.4pt}{-0.4pt}}
    \pgfpathsine{\pgfpoint{0.4pt}{-0.4pt}}
    \pgfpathcosine{\pgfpoint{0.4pt}{0.4pt}}
    \pgfusepath{stroke}
  },
  transform shape
}}

\usepackage[hyphens]{url}
\usepackage{hyperref}
\hypersetup{breaklinks, colorlinks, linkcolor=MidnightBlue, anchorcolor=MidnightBlue, citecolor=MidnightBlue, urlcolor=MidnightBlue}

\usepackage[ruled,vlined,linesnumbered]{algorithm2e}
\SetKwInput{kwInit}{Initialization}

\input{header}

\newcommand{\ED}{{\sf ED}}
\newcommand{\SCED}{\text{SCED}}
\newcommand{\PSCED}{\text{P-SCED}}
\newcommand{\CSCED}{\text{C-SCED}}
\newcommand{\RSCED}{\text{R-SCED}}

\newcommand{\LMPmar}{\text{S-LMP}}
\newcommand{\LMPnom}{\text{N-LMP}}

\newcommand{\N}{{\sf N}}
\renewcommand{\S}{{\sf S}}

\newcommand{\DA}{{\sf DA}}
\newcommand{\SE}{{\sf SE}}

\newcommand{\voll}{{\v{v}}}
\newcommand{\VoLL}{\text{VoLL}}

\newcommand{\CVaR}{{\textrm{CVaR}}}
\newcommand{\cvar}[2][]{{\CVaR{}\ifx\relax\detokenize{#1}\relax\else_{#1}\fi{\left[#2\right]}}}

\renewcommand{\bone}{\mathds{1}}

\newcommand{\MS}{{\rm MS}}

\newcommand{\LOC}{{\rm LOC}}

\DeclareMathOperator{\sign}{sign}

\title{Risk-Sensitive Security-Constrained Economic Dispatch: Pricing and Algorithm Design}
\author{Avinash N. Madavan\quad Nathan Dahlin \quad Subhonmesh Bose \quad Lang Tong\thanks{A. N. Madavan and S. Bose are with the Department of Electrical and Computer Engineering at the University of Illinois Urbana-Champaign. N Dahlin is with the Department of Electrical and Computer Engineering at the University at Albany, SUNY. L. Tong is with the School of Electrical and Computer Engineering at Cornell University. Emails: {avinash.madavan@gmail.com, ndahlin@albany.edu, boses@illinois.edu, lt35@cornell.edu}. This work was partially supported by a grant from the National Science Foundation under the grant NSF CAREER 2048065, ISO New England, and the Power Systems Engineering Research Center.}}
\date{}

\newcommand{\rev}[1]{{\textcolor{black}{{#1}}}}
\newcommand{\revision}[1]{{\textcolor{black}{{#1}}}}

\renewcommand{\hat}{\widehat}

\usepackage[normalem]{ulem}

\usepackage[condensed,math]{kurier}
\usepackage[T1]{fontenc}

\begin{document}

\maketitle
%
\input{abstract}
%
\input{introduction}
%
\input{ed}
%
\input{rsced}
\input{pricing}

\input{algorithm}
%
\input{simulation}
\input{conclusion}

\bibliographystyle{ieeetr}
\bibliography{rsced_bib}
\appendix
\input{proofs}
\end{document}

%% file: header.tex
\newcommand{\beq}{\begin{equation}}
\newcommand{\eeq}{\end{equation}}
\newcommand{\beqa}{\begin{eqnarray}}
\newcommand{\eeqa}{\end{eqnarray}}
\newcommand{\beqan}{\begin{eqnarray*}}
\newcommand{\eeqan}{\end{eqnarray*}}

\newcommand{\E}{\mathds{E} }

\newcommand{\Rset}{\mathbb{R}}

\newcommand{\Xset}{\mathbb{X}}

\newcommand{\Ccal}{{\cal C}}

\newcommand{\Gcal}{{\cal G}}

\newcommand{\bone}{\mathbf{1}}

\renewcommand{\v}[1]{{\bm{#1}}}

\newcommand{\ol}[1]{\ensuremath{\overline{{#1}}}}
\newcommand{\ul}[1]{\ensuremath{\underline{{#1}}}}

\renewcommand{\]}{\right]}

\newcounter{l1}
\newcounter{l2}
\newcounter{l3}
\newcommand{\bdotlist}{\begin{list}{$\bullet$}{}}
\newcommand{\bboxlist}{\begin{list}{$\Box$}{}}
\newcommand{\bbboxlist}{\begin{list}{\raisebox{.005in}{{\tiny
$\blacksquare$ \ \ }}}{}}
\newcommand{\bdashlist}{\begin{list}{$-$}{} }
\newcommand{\blist}{\begin{list}{}{} }
\newcommand{\barablist}{\begin{list}{\arabic{l1}}{\usecounter{l1}}}
\newcommand{\balphlist}{\begin{list}{(\alph{l2})}{\usecounter{l2}}}
\newcommand{\bAlphlist}{\begin{list}{\Alph{l2}.}{\usecounter{l2}}}
\newcommand{\bdiamlist}{\begin{list}{$\diamond$}{}}
\newcommand{\bromalist}{\begin{list}{(\roman{l3})}{\usecounter{l3}}}

\newtheorem{theorem}{Theorem}

\newtheorem{definition}{Definition}

%% file: abstract.tex

\begin{abstract}
	We propose a risk-sensitive security-constrained economic dispatch (\RSCED{}) formulation capturing the tradeoff between dispatch cost and resilience against potential line failures,
    where risk is modeled via the conditional value at risk (\CVaR{}). In the context of our formulation, we analyze revenue adequacy and side payments of two pricing models, one based on nominal generation costs, and another based on total marginal cost including contingencies. In particular, we prove that the system operator's (SO) merchandising surplus (MS) and total revenue are nonnegative under the latter, while under the former the same does not hold in general. 
    We demonstrate that the proposed \RSCED{} formulation is amenable to decomposition and describe a Benders' decomposition algorithm to solve it. In numerical examples, we illustrate the differences in MS and total revenue under the considered pricing schemes, and the computational efficiency of our decomposition approach.  
\end{abstract}

%% file: introduction.tex

\section{Introduction}

Power procurement is typically performed by solving an \emph{economic dispatch} (ED) problem matching aggregate supply to demand and ensuring an economically efficient dispatch satisfying network constraints. These problems are often solved in forward planning operations prior to the time of power delivery. Such forward planning faces uncertainty in many aspects including available supply, demand fluctuations, and component availability. In this paper, we focus on uncertainty in transmission line availability. In order to ensure that system operators (SOs) can respond adequately to any line failures, they consider ED problems augmented with additional security constraints known as \emph{security-constrained economic dispatch} (\SCED{}) problems. Various \SCED{} formulations exist, many of which fall under two primary categories: preventive security (\PSCED{}), which ensures dispatches satisfy line flow capacities across considered contingencies \cite{alsac1974optimal}, and corrective security (\CSCED{}), which allows for an SO to take corrective action in the form of reserve capacity deployment \cite{monticelli1987security,capitanescu2007improving,li2016adaptive,capitanescu2011state} and in some cases load shed \cite{bouffard2008stochastic}.

Reserve capacity is itself procured in forward planning operations. Historically, this procurement has been guided by heuristics, e.g., securing reserves up to a fixed percentage of the total anticipated load as in the case of CAISO \cite{CAISO}. However, such methods 
can result in either larger or smaller than necessary quantities of procured capacity. The former results in additional overhead costs and inefficient dispatches, while the latter can result in involuntary blackouts. Here we consider a \SCED{} formulation explicitly incorporating and thus automating the reserve procurement process, 
similar to the approaches in \cite{capitanescu2011state,shi2022scenario,galiana2005scheduling}. Specifically, we simultaneously optimize over-dispatch and corrective actions, yielding explicit quantities of reserve capacity required across failure scenarios as a byproduct. Therefore, sufficient reserve capacity is procured while minimizing the cost associated with this capacity.

In the context of power system operation, security can be defined in either a deterministic or probabilistic sense \cite{galiana2005scheduling}. Deterministic definitions require that demand is met without load shed across all contingencies, while probabilistic definitions allow that scenario-dependent portions of load may be shed at an additional cost to be minimized. Traditionally this additional cost has been modeled as the expected \emph{value of lost load} (\VoLL{}) with respect to a reference probability distribution over failure scenarios, e.g., as in \cite{bouffard2008stochastic}. However, SOs usually treat load shed as a measure of last resort, and optimization of expected \VoLL{} can in practice expose power systems to undesirably high levels of risk for load shed. On the other hand, robust approaches based on deterministic notions of security tend to result in overly conservative dispatches suffering similar overhead costs as in some of the aforementioned heuristic approaches. Thus, a fundamental tradeoff exists between dispatch reliability in the face of uncertainty and economic efficiency, and existing \SCED{} formulations tend to implicitly impose a choice between one or the other. 

In this work, we propose a \emph{risk-sensitive} \SCED{} (\RSCED{}) formulation which features simultaneous primary and reserve generation, as well as a simple parametrization capable of capturing a continuum of risk preferences towards load shed. 
In particular, we model the risk associated with load shed costs via the \emph{conditional value at risk} (\CVaR{}) measure. This risk measure, parameterized by a single scalar $\alpha$, measures the average losses over the $1 - \alpha$ fraction of worst-case scenarios. Adjusting $\alpha$ allows SOs to express their tolerance towards potential load shed. For example, setting $\alpha = 0$ recovers consideration of expected cost over all scenarios, while setting $\alpha$ close to 1 focuses optimization on worst-case, i.e., highest-cost scenarios, approaching deterministically robust solutions.

The \CVaR{} measure has garnered recent attention in power system applications \cite{jabr2005robust,asensio2015stochastic,madavan2019risk,li2018flexible,wang2015distributionally} due to several properties that make it highly amenable to optimization. Most notably, 
\CVaR{} is a coherent risk measure, which implies that it preserves convexity \cite{shapiro2021lectures} of input cost functions, and as a special case retains the linearity of input linear cost structures. 
As such, optimization of \CVaR{} objectives allows for the use of off-the-shelf convex optimization solvers and adoption of decomposition methods such as Benders' decomposition that have been applied to similar problems in \cite{liu2015computational}.

The convexity of \CVaR{} further aids in establishing meaningful prices for the \RSCED{}-based dispatches via Lagrangian analysis. SO revenue adequacy and individual rationality for participating generators are crucial prerequisites to a functioning market. In the context of the proposed \RSCED{} formulation, we examine whether these properties can be guaranteed under two deterministic market clearing mechanisms suggested for use in concert with \SCED{} formulations. The first,
which we refer to as the \emph{locational marginal price of the nominal} (\LMPnom{}), establishes prices based solely on the nominal dispatch, i.e., the base case dispatch with respect to which potential corrective actions are taken. Proponents of these prices observe that explicitly pricing against security constraints results in increased cost due to unrealized scenarios \cite{boucher1998security}. In contrast, authors in \cite{hogan2014electricity} suggest that these prices are inadequate to capture the cost borne by SOs or support grid maintenance and development, and instead advocate for the use of an ex-ante pricing scheme that we refer to as the \emph{locational marginal price under security} (\LMPmar{}). These prices reflect the total marginal cost of generation in \CSCED{} problems, including our \RSCED{} formulation.

Prior work on pricing design for \SCED{} problems includes a variety of results on both revenue adequacy and cost recovery. Note that cost recovery is implied by individual rationality. Considering energy-only formulations and two-stage, stochastic pricing mechanisms \cite{pritchard2010single} and \cite{morales2012pricing} establish both properties in expectation, while \cite{morales2014electricity} establishes cost recovery in expectation and revenue adequacy in all scenarios. In \cite{wong2007pricing}, an energy-reserve co-optimization formulation paired with \LMPmar{} pricing is likewise shown to yield both properties in expectation. 
Incorporating separate costs associated with reserve deployment in an energy-reserve formulation, authors in \cite{shi2022scenario} demonstrated revenue-adequacy in expectation and cost recovery per scenario. Our risk-sensitive \CVaR{}-based formulation generalizes these former ones, which optimize expected costs. \rev{While risk-sensitive dispatch formulations, as well as \LMPnom{} and \LMPmar{} pricing mechanisms, have been proposed separately in the literature, the market-relevant properties such as revenue adequacy and individual rationality of these pricing mechanisms remain poorly understood, including in risk-sensitive contexts.}

The main contributions of this paper are as follows:
    
    \noindent $\bullet$ A scenario-based \emph{risk-aware} energy-reserve co-optimization model is proposed. Risk is modeled via the \CVaR{} risk measure, resulting in a convex optimization problem incorporating network constraints in all contingency scenarios. 
    
    \noindent $\bullet$ \rev{Individual rationality is established for \LMPnom{} and \LMPmar{} variants of locational marginal prices derived in our risk-aware setting, when paired with associated \emph{lost opportunity cost} (\LOC{}) payments made by the SO to generators 
    for deviation from individual profit-maximizing dispatches.}
    
    \noindent $\bullet$ Revenue adequacy is established for  \LMPmar{} pricing of \RSCED{}, while we show that \LMPnom{} may result in negative revenue. Including \LOC{} payments, we further show that the total revenue under \LMPmar{} pricing is nonnegative. 

    \noindent $\bullet$ We provide a Benders' decomposition algorithm to efficiently solve the \RSCED{} problem in a manner analogous to \rev{risk-neutral or robust SCED approaches}. 

\revision{Our proposed \RSCED{} formulation can be used within existing electricity market operations in a variety of contexts. For example, one can use \RSCED{} to clear the 5-15 minute ahead real-time market, for which our analyzed pricing schemes define the  settlements for the market participants. \RSCED{} can also be used within a residual unit commitment module that SOs often run an hour to several hours in advance of power delivery. An enhanced multi-period version of the same with inter-temporal constraints can be used within a security-constrained unit commitment and economic dispatch module to clear the day-ahead market within organized wholesale market structures. One can theoretically use our framework to address long-term planning or conceive a capacity market design with additional work. However, as it will become clear, the size of \RSCED{} grows with the number of scenarios considered, which can be prohibitively large to capture operating points across planning horizons.}

\revision{In this paper, we concretely focus on single line failures as possible contingencies throughout. Our design principle and the formulation technically does not require that we only model single line failures. A power system in practice has a large number of components; one can create a contingency scenario by considering the failure of any possible subset among these components, and construct a finite collection of such scenarios. Our design philosophy will apply to this variant, where the optimization problem will seek to compute a nominal dispatch and reserve procurement in a way that penalizes the risk-based cost of said procurement and possible recourse actions within these scenarios, given operational constraints.}

\revision{In our analysis of the settlements, we narrowly focus on market-relevant properties such as individual rationality and revenue-adequacy. Organized wholesale market settlements are complex; their payments can depend on considerations that we do not explicitly model. However, our analysis is premised on the belief that, if a market settlement scheme, sans those complications, is theoretically favored to another based on the market-relevant properties we analyze, then we expect that design to fare better in practice with additional considerations. As will become evident, \LMPmar{} outperforms \LMPnom{} in the metrics we consider, and hence, defines a more sound pricing scheme from a theoretical standpoint. Said succinctly, \LMPmar{} explicitly accounts for the marginal costs of recourse actions in contingencies, while \LMPnom{} ignores them. As the power system grows in complexity, and the range of uncertainty increases, ignoring components due to responses to these uncertain scenarios in prices offered to the generators might not reflect the true value of these resources within the system. 
}

The paper is organized as follows. We begin by presenting established results on the deterministic economic dispatch problem, LMP-based market design, and properties of such markets in Section \ref{sec:ED}. We then formulate the \RSCED{} problem in Section \ref{sec:rsced}, contextualizing it within existing work. In Section \ref{sec:pricing}, we study the market design under \LMPnom{} and \LMPmar{}, providing rigorous proofs of the revenue achieved in clearing the market. We present, in Section \ref{sec:algorithm}, a Benders' decomposition algorithm to solve \RSCED{} efficiently at scale. Finally, Section \ref{sec:simulation_results} describes simulation results of the \RSCED{} formulation on the IEEE 24-bus RTS network, with a scalability demonstration of the Benders' decomposition approach for a large array of networks. \revision{We introduce notation where used, but key symbols are included in Table \ref{tab:symbols}.}

{\small
\begin{table}[htbp]
\caption{Major symbols}\label{tab:symbols}
\begin{center}
\vspace{-1em}
\begin{tabular}{ll}
\hline
$\v{g}$& Nominal nodal generation dispatch\\
$\ul{\v{r}},\,\ol{\v{r}}$& Negative and positive nodal reserve capacity\\
$\v{\delta g}_k$&Nodal generator re-dispatch in contingency $k$\\
$\v{\delta d}_k$&Nodal load shed in contingency $k$\\
$\alpha$&Tunable risk parameter\\
$\MS{}$& Merchandising surplus\\
$\LOC{}$& Lost opportunity cost\\
$\v{v}$&Value of lost load (VoLL)\\
$\v{\pi}^{\N}$&Locational marginal prices of the nominal\\
$\v{\pi}^{\S}$&Locational marginal prices under security\\
\hline
\end{tabular}
\end{center}
\end{table}
}

%% file: ed.tex
\section{Preliminaries}
\label{sec:ED}

SOs are tasked with making a series of forward dispatch decisions prior to real-time demand fulfillment. 
They do so by solving \emph{economic dispatch} (ED) problems, which seek to minimize the cost of energy procurement, or dispatch cost, subject to engineering constraints of the grid \cite{varaiya2010smart}. We begin by studying the deterministic economic dispatch problem, without modeling potential component failures or recourse action, in order to identify desirable properties characterizing its outcomes to be extended to its risk-sensitive variants. We refer to the works in \cite{varaiya2010smart,ma2009security,rajagopal2013risk} for further details on existing economic dispatch formulations and practices. 

Consider an $n$-bus power network, connected by $\ell$ lines. Throughout the paper, for a positive integer $m$, we use $[m]:=\{1,\dots,m\}$. 
Located at each bus are both demand and generation, the vector values of which are denoted by $\v{d} \in \mathbb{R}^n$ and $\v{g} \in \mathbb{R}^n$. Assume that generator $i$ has a linear cost $c_i$, collected by the vector $\v{c}$,
and that power production is limited to the set $\Gcal := [\underline{\v{g}}, \overline{\v{g}}]$. The total cost of generation is then given by $\v{c}^\top \v{g}$. We note that piecewise affine costs may be adopted as well without introducing additional conceptual difficulties. Let $\v{f} \in \mathbb{R}^{2\ell}$ denote the vector of (directed) line flow limits. We adopt the DC power flow approximation, which assumes lossless lines, negligible reactive power, and small voltage phase angle differences. Such assumptions allow us to represent the nonlinear constraints dictated by  Kirchhoff's laws as linear constraints. Specifically, let $\v{H} \in \mathbb{R}^{2 \ell \times n}$ be the injection-shift factor (ISF) matrix mapping the injections to (directed) line flows, and $\bone$ denote the vector of all ones of appropriate size. Then, the ED problem takes the form
\begin{subequations}
\begin{align}
    \text{(ED)}\,\,\,&\underset{\v{g} \in \Gcal}{\text{minimize}} && \v{c}^\top \v{g}, \label{eq:ED.obj} \\
    &\text{subject to} && \bone^\top \left(\v{g} - \v{d}\right) = 0, \quad \v{H} \left( \v{g} - \v{d} \right) \le \v{f}. \label{eq:ED.cons}
\end{align}
\label{eq:ED}
\end{subequations}
In order to design a compensation structure,
assign dual multipliers $\lambda \in \mathbb{R}$ and $\v{\mu} \in \mathbb{R}^{2\ell}$ to each of the constraints in \eqref{eq:ED.cons}. Let $(\v{g}^\star, \lambda^\star, \v{\mu}^\star)$ be an optimal primal-dual solution to the ED problem \eqref{eq:ED}. Then the \emph{nodal prices} or \emph{locational marginal prices} (LMPs) are defined as,
\begin{equation}\label{eq:ED_LMP}
    \v{\pi}^{\ED} := \lambda^\star \bone - \v{H}^\top \v{\mu}^\star.
\end{equation}
These prices are known to satisfy a number of useful properties. Notably, this LMP is the marginal sensitivity of the optimal cost to demand. That is, a unit increase in demand at node $i$ would result in a corresponding increase in the total procurement cost of magnitude $\pi^{\ED}_i$.

In clearing the market, SOs accept payments $\pi_i d_i$ from each consumer $i\in[n]$ and pay $\pi_ig^{\star}_i$ to each supplier $i\in[n]$ for the quantities procured and provided, respectively. 
The summed consumer payments need not equal the summed payments made to suppliers, i.e., SOs may operate at a profit or loss. This difference is captured by the \emph{merchandising surplus} (\MS{}). 
Given nodal prices $\v{\pi}^{\ED}$, the merchandising surplus is 
\begin{equation}\label{eq:MS}
    \MS{}[\v{\pi}^{\ED}] := (\v{\pi}^{\ED})^\top (\v{d} - \v{g}^\star),
\end{equation}
It can be shown that $\MS{}[\v{\pi}^{\ED}] \ge 0$. 
Such a market is called \emph{revenue adequate}, which guarantees that SOs do not run cash-negative from operating a market. Moreover, in the case of strictly positive merchandising surplus, additional revenue is disseminated through financial derivatives such as financial transmission rights. 

Assuming price-taking behavior amongst market participants, LMP-based market clearing satisfies the condition of \emph{individual rationality}. That is, for any $i\in[n]$ and corresponding LMP $\pi^{\ED}_i$, the SO prescribed dispatch $g^\star_i$ maximizes the profit of generator $i$, or
\begin{equation}\nonumber
    g_i^\star \in \underset{\hat{g}_i \in \mathcal{G}_i}{\arg\max} \; (\pi^{\ED}_i - c_i) \hat{g}_i,
\end{equation}
where $\mathcal{G}_i$ denotes the interval $[\ul{g}_i,\ol{g}_i]$. Thus, under individually rational prices, generators have no incentive to deviate from the prescribed dispatch. Failing to enforce this property obligates SOs to compensate generators for any differences between the prescribed and profit-maximizing dispatch. Such a concern arises elsewhere in power system operations, e.g., in the unit commitment problem, where generators are paid \emph{uplift payments} in compensation for the costs associated with startup and shutdown \cite{yang2019unified}. Payments of this sort are problematic in that they are out-of-market settlements, whose costs are not reflected in objective functions representing cost minimization or social welfare maximization.

%% file: rsced.tex

\section{Formulating the risk-sensitive security-constrained economic dispatch problem}
\label{sec:rsced}

ED problem \eqref{eq:ED} assumes a static network topology, without regard for possible component failures. To ensure resilience against potential outages, this formulation must be augmented with additional security constraints. 


We now present our formulation of the risk-sensitive counterpart of ED problem \eqref{eq:ED}. Consider a set of credible line outage scenarios, indexed by $k\in[K]$, in which one or more lines fail. Suppose scenario $k$ is realized\footnote{\rev{Line outage occurance probabilities, as well as average outage durations, can be obtained from historical data, e.g, by methods described in \cite{chowdhury2006causal}.}}. Then, the post-contingency line flows are given by $\v{H}_k (\v{g} - \v{d})$, where $\v{H}_k$ is the ISF matrix describing the post-contingency network. Such flows must remain within rated line capacities. However, these capacities arise primarily from thermal considerations and can be relaxed temporarily. We refer to the situation in which no contingency $k\in[K]$ arises as the nominal or the base-case. 
We adopt the convention in \cite{national2010transmission}, summarized in Table \ref{tab:relaxed.line.limits}, allowing capacities to be relaxed to the drastic-action limit $\v{f}_k^\DA{}$, in the event of a failure. We ensure flows return to the short-term emergency level $\v{f}_k^\SE{}$, within 5 minutes, during which we allow for some recourse action. \footnote{\rev{In this work, we allow for an initial dispatch decision, followed by a single redispatch once a scenario occurs. Our framework is readily extended to multi-period settings, where a series of redispatches are taken to bring line flows within normal operating limits.}}
\begin{table}[ht]
    \centering
    \caption{Relaxed line flow capacity limits for emergency scenarios.}
    \begin{tabular}{l c c}
        \toprule
         \textbf{Name} & \textbf{Maximum allowable time} \\
         \midrule
         Nominal & Indefinite \\
         Short-Term Emergency (SE) & 15 minutes \\
         Drastic Action (DA) & 5 minutes \\
         \bottomrule \\
    \end{tabular}
    \label{tab:relaxed.line.limits}
\end{table}
%
We assume that the recourse actions available to SOs include re-dispatch of generation and load shed. By simultaneously procuring generation and reserves, we seek to ensure that adequate capacity is available in line outage scenarios, and reduce the occurrence and magnitude of load shed. 

Let $\underline{\v{r}}$ and $\overline{\v{r}}$ denote the vectors of negative and positive nodal reserve capacity, with associated linear marginal costs $\underline{\v{c}}_r$ and $\overline{\v{c}}_r$, respectively. Collecting generator re-dispatch
and load shed decisions across scenarios $[K]$ as $\v{\delta g}$ and $\v{\delta d}$, respectively, and 
defining the acceptable load shed range in any scenario as $\mathcal{D} := [0, \v{\Delta}_d]$, the risk-sensitive economic dispatch problem $\RSCED{}_{\alpha}$ is given by
\begin{subequations} %
\label{eq:RSCED} %
\begin{align} %
    &\underset{\v{g}, \underline{\v{r}}, \overline{\v{r}}, \v{\delta g}, \v{\delta d}}{\text{minimize}} && \v{c}^\top \v{g} + \underline{\v{c}}_r^\top \underline{\v{r}} + \overline{\v{c}}_r^\top \overline{\v{r}} + \CVaR{}_{\alpha} \left[ \mathcal{C}(\v{\delta d} ) \right] \label{eq:RSCED.obj}, \\
    &\text{subject to} && \bone^\top \left(\v{g} - \v{d}\right) = 0, \; \v{H} \left( \v{g} - \v{d} \right) \le \v{f}, \; \v{g} \in \Gcal \label{eq:RSCED.nominal} \\
    &&& \v{H}_k \left( \v{g} - \v{d} \right) \le \v{f}_k^{\DA},\label{eq:RSCED.da} \\
    &&& \bone^\top \left( \v{\delta g}_k + \v{\delta d}_k \right) = 0,\label{eq:RSCED.se.balance} \\
    &&& \v{H}_k \left( \v{g} + \v{\delta g}_k - \v{d} + \v{\delta d}_k \right) \le \v{f}_k^{\SE}, \label{eq:RSCED.se.flow} \\
    &&& \v{g} + \v{\delta g}_k \in \Gcal, \; \v{\delta g}_k \in [-\underline{\v{r}}, \overline{\v{r}}], \label{eq:RSCED.se.genbounds} \\
    &&& 0 \le \underline{\v{r}}, \overline{\v{r}}, \; \v{\delta d}_k \in \mathcal{D}, \label{eq:RSCED.se.bounds} \\
    \nonumber &&& \forall k\in [K]
\end{align} %
\end{subequations} %
where in \eqref{eq:RSCED.obj}, $\Ccal(\v{\delta d})$ denotes the random load shed cost given $\v{\delta d}$. The \emph{conditional value at risk} measure $\CVaR{}_{\alpha}$ and associated parameter $\alpha$ are formally defined below. 

The cost associated with load shed is modeled by the \emph{value of lost load} (\VoLL{}), denoted $\voll\in\Rset^n$. The realized \VoLL{} varies over the line outage scenarios $[K]$. Assuming that contingencies are independent, \rev{i.e., only one contingency occurs at a time,} and each contingency $k$ occurs with probability $p_k$, the random load shed cost is given by
\begin{equation}
    \mathcal{C}(\v{\delta d}) := \begin{cases}
      \voll^\top \v{\delta d}_k & \text{with probability $p_k$ for $k\in[K]$}, \\
      0 & \text{with probability } 1 - \sum_{k = 1}^K p_k.
    \end{cases}
    \label{eq:voll}
\end{equation}

\begin{figure}
    \centering
    \includegraphics[width=0.6\linewidth]{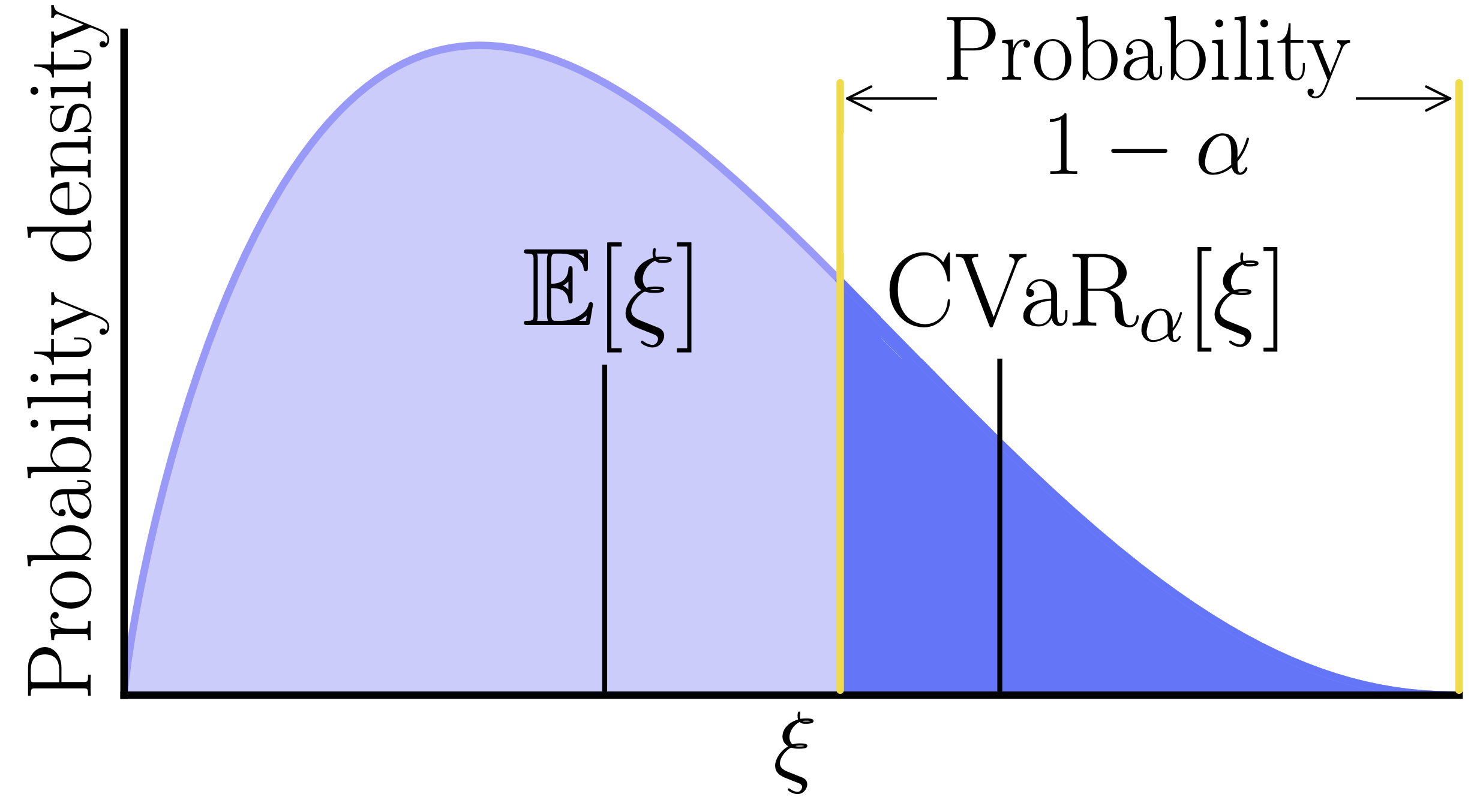}
    \caption{A description of \CVaR{} over the cost distribution induced by a set of fixed decision variables}
    \label{fig:cvar}
\end{figure}

Together with generation and reserve capacity costs, under the \RSCED{} formulation in \eqref{eq:RSCED} we minimize the conditional value at risk (\CVaR{}) associated with $\Ccal(\v{\delta d})$. Intuitively, given parameter $\alpha$, $\CVaR{}_{\alpha}$
measures the expected tail loss over the $(1-\alpha)$-fraction of worst-case scenarios. For general distributions over scaler random variable $\xi$, $\CVaR{}_{\alpha}[\xi]$ is formally defined via the variational form
\begin{equation}\label{eq:CVaR_var}
    \cvar[\alpha]{\xi} := \min_{\mathpzc{z}} z + \frac{1}{1 - \alpha} \mathbb{E}[ \xi - z ]^+,
\end{equation}
where $[\cdot]^+$ denotes the positive part of its argument  \cite{rockafellar2000optimization}, and the expectation is taken with respect to the probabilities $p_k$ for $k\in[K]$ appearing in \eqref{eq:voll}. Selection of $\alpha$ allows an SO to express their level of tolerance to high cost scenarios. Setting $\alpha = 0$ renders \CVaR{} equal to expected value, while taking $\alpha \uparrow 1$, drives \CVaR{} to the essential supremum, akin to robust optimization. Figure \ref{fig:cvar} provides an illustration of this risk-measure over a fixed distribution.
Optimization over a \CVaR{} objective amounts to selecting decision variables whose induced distribution reduces the expected tail loss. 

SOs typically treat load shed as an action of last resort and are thus highly averse to such outcomes, and minimization of expected costs can lead to undesirably high levels of load shed. Instead, the use of \CVaR{} takes advantage of the fact that \VoLL{} is typically much larger than the cost of generation or reserves.
Thus, increasing the risk aversion parameter $\alpha$ 
prioritizes the avoidance of load shed.

\revision{Specific transmission lines may be more prone to failures than others, e.g., from wildfire risks as modeled in \cite{rhodes2020balancing,taylor2022framework}. We allow these line failure probabilities $p_1, \ldots, p_K$ to vary across the scenarios and study the impact of such probabilities empirically later in the paper. One can consider multi-line and generator failures within our formulation. In this work, we focus on the case of single line failures to streamline our presentation. When including possible generator failures, one needs to appropriately constrain recourse actions that can be taken in each such contingency.}

\subsection{Formulating \RSCED{} as a linear program}\label{sec:formulating}
Given the piecewise linearity of $\Ccal(\v{\delta d})$ across scenarios, the \RSCED{} problem can be written as a large linear program. 
Applying \eqref{eq:CVaR_var} to $\CVaR_{\alpha}[\Ccal(\v{\delta d})]$, the \RSCED{} objective \eqref{eq:RSCED.obj} can equivalently be written as
\begin{equation}
\label{eq:RSCED_obj_var}
\begin{aligned}
    & \min_z \; \v{c}^\top \v{g} + \underline{\v{c}}_r^\top \underline{\v{r}} + \overline{\v{c}}_r^\top \overline{\v{r}} + z + \frac{1}{1 - \alpha} \E[\mathcal{C}(\v{\delta d}) - z]^+ \\
    &  = \min_z \; \v{c}^\top \v{g} + \underline{\v{c}}_r^\top \underline{\v{r}} + \overline{\v{c}}_r^\top \overline{\v{r}} + z + \frac{1}{1 - \alpha} \sum_{k = 1}^K p_k [\voll^\top \v{\delta d}_k - z]^+.
\end{aligned}
\end{equation}
Taking the epigraph form of each term in the sum over scenarios in \eqref{eq:RSCED_obj_var}, the \RSCED{} problem can be written as
\begin{subequations}
\label{eq:RSCED.lp}
\begin{flalign}
    &\underset{\substack{z, \v{y}, \v{g}, \underline{\v{r}}, \overline{\v{r}}, \\ \v{\delta g}, \v{\delta d}}}{\text{minimize}} && \v{c}^\top \v{g} + \underline{\v{c}}_r^\top \underline{\v{r}} + \overline{\v{c}}_r^\top \overline{\v{r}} + z + \frac{1}{1 - \alpha} \sum_{k = 1}^K p_k y_k, \\ 
    &\text{subject to} && (\lambda, \underline{\v{\gamma}}, \overline{\v{\gamma}}) : \bone^\top \left(\v{g} - \v{d}\right) = 0, \; \v{g} \in \Gcal \label{eq:RSCED.lp.nominal} \\
    &&& (\v{\mu}): \v{H} \left( \v{g} - \v{d} \right) \le \v{f},  \label{eq:RSCED.lp.lineflow}\\
    &&& (\v{\mu}_k^\DA{}): \v{H}_k \left( \v{g} - \v{d} \right) \le \v{f}_k^{\DA},  \label{eq:RSCED.lp.lineflow_k}\\
   &&& (\lambda_k): \bone^\top \left( \v{\delta g}_k + \v{\delta d}_k \right) = 0, \label{eq:RSCED.lp.se.balance} \\
    &&& (\v{\mu}_k^\SE{}): \v{H}_k \left( \v{g} + \v{\delta g}_k - \v{d} + \v{\delta d}_k \right) \le \v{f}_k^{\SE}, \label{eq:RSCED.lp.se.flow} \\
    &&& (\underline{\v{\gamma}}_k, \overline{\v{\gamma}}_k): \v{g} + \v{\delta g}_k \in \Gcal, \label{eq:RSCED.lp.se.genbounds} \\
    &&& (\underline{\v{\rho}}_k, \overline{\v{\rho}}_k): \v{\delta g}_k \in [-\underline{\v{r}}, \overline{\v{r}}], \label{eq:RSCED.lp.se.dgenbounds} \\
    &&& (\underline{\v{\eta}}, \overline{\v{\eta}}, \underline{\v{\sigma}}_k, \overline{\v{\sigma}}_k): 0 \le \underline{\v{r}}, \overline{\v{r}}, \; \v{\delta d}_k \in \mathcal{D}, \label{eq:RSCED.lp.se.bounds} \\
    &&& (\underline{\nu}_k, \overline{\nu}_k): y_k \ge 0, \; y_k \ge \voll^\top \v{\delta d}_k - z, \label{eq::RSCED.lp.epigraph}\\ 
    \nonumber&&& \forall k\in[K],
\end{flalign}
\end{subequations}
where we have associated Lagrange multipliers (in parentheses) with each of the constraints. As with the LMP-based prices arising from the ED problem \eqref{eq:ED}, these dual multipliers will constitute the basis for the R-SCED based market clearing process and prove useful in establishing its properties.

\revision{An important implication of the formulation of R-SCED as a linear program in \eqref{eq:RSCED.lp} is that it falls under the same optimization class as previously studied C-SCED formulations presented in \eqref{eq:csced} below. Despite adding the flexibility of specifying risk-aversion through the tunable parameter $\alpha$ (the effect of which we will study through examples), R-SCED lies in the same computational class as does C-SCED.}

\subsection{Relation to existing formulations}
Various security-constrained economic dispatch formulations exist within the literature, offering different guarantees for security. Perhaps the simplest is \emph{preventive}-\SCED{} (\PSCED{}) \cite{alsac1974optimal}, which is given by
\begin{equation}\nonumber
\begin{aligned}
(\PSCED{})\quad& \underset{\v{g}}{\text{minimize}} && \v{c}^\top \v{g}, \\
    & \text{subject to} && \eqref{eq:RSCED.nominal}, \v{H}_k (\v{g} - \v{d}) \le \v{f},\\
    &&& \forall k\in[K]
\end{aligned}
\end{equation}
\PSCED{} ensures that the nominal system dispatch is robust to \emph{any} credible outage, but can be overly conservative. Corrective-\SCED{} (\CSCED{}) \cite{monticelli1987security} allows for potential recourse action, but typically assumes fixed reserve capacity, 
ignores the cost associated with recourse action, and does not allow for load shed (i.e., $\v{\delta d}_k$ is fixed at $\v{0}$ in all scenarios)
\begin{equation}\label{eq:csced}
\begin{aligned}
    (\CSCED{})\quad&\underset{\v{g}, \v{\delta g}}{\text{minimize}} && \v{c}^\top \v{g}, \\
    & \text{subject to} && \eqref{eq:RSCED.nominal} \text{-} \eqref{eq:RSCED.se.balance}, \eqref{eq:RSCED.se.genbounds}, \\
    &&& \v{H}_k (\v{g} + \v{\delta g}_k - \v{d}) \le \v{f}_k^\SE{},\\
     &&& \forall k \in [K].%
\end{aligned}
\end{equation}
To illustrate the differences in dispatch and costs rendered under ED, \PSCED{}, and \CSCED{}, consider the 3-bus network in \cite{lesieutre2011examining}  with added generation at bus 3, shown in Figure \ref{fig:3bus}. 
\input{3bus}

In this example, all generation capacity limits are $20$ MW and lines 1-2 and 1-3
have capacity $9000$ MW, and thus can be neglected. Line 2-3 has a capacity of $50$ MW. Assume drastic action limits and short term emergency limits 
given by $\v{f}_k^\DA{} = 1.8 \v{f}$ and $\v{f}_k^\SE{} = 1.2 \v{f}$, respectively, and the maximum reserve procurement to be $0.2$ MW per generator. 
%
\begin{table}[ht]
    \centering
        \caption{Comparison of dispatch for various ED formulations.}
    \begin{tabular}{l c c c}
         \toprule
         \textbf{Method} & ($g_1^*$, $g_2^*$, $g_3^*$) \textbf{(MW/h)} & \textbf{Nominal cost (\$/h)}  \\
         \midrule
         ED & (144.3, 170.7, 0) & 926 \\
         \PSCED{} & (110, 160, 45) & 1192 \\
         \CSCED{} & (119, 181, 15) & 962  \\
         \bottomrule \\
    \end{tabular}
    \label{tab:ED.compare}
\end{table}
Comparing nominal cost figures in Table \ref{tab:ED.compare}, we observe that \CSCED{} produces a lower cost of generation than \PSCED{}, reflecting the cost-free adjustments allowed for in \eqref{eq:csced}. While the ED dispatch achieves the lowest cost of the three methods, this lower cost comes at the expense of any sort of security guarantee against line failure contingencies. 

We now consider \RSCED{} outcomes including reserve procurement and load shed for the 3-bus network. The cost of reserve capacity is given by $1.2$ times the cost of generation, and the value of lost load is uniformly $\$30$/MWh. Assume that the probability of a failure of any single line is $p_k = 0.1$. 
%
%
\begin{table}[H]
    \centering
        \caption{Comparison of \RSCED{} dispatch for various $\alpha$.}\label{tab:RSCED.compare}
    \begin{tabular}{l c c c c}
         \toprule
         & $\v{g}^*$ & \textbf{Nominal} & \textbf{Reserve} & \textbf{Total load} \\
         \textbf{Method} & \textbf{(MW/h)} & \textbf{cost (\$/h)} & \textbf{cost (\$/h)} & \textbf{shed (MW)}  \\
         \midrule
         \RSCED{}$_0$ & (119, 181, 15) & 962 & 28.8 & 31 \\
         \RSCED{}$_{0.1}$ & (110, 184.7, 20.3) & 974.9 & 21.1 & 29.34 \\
         \RSCED{}$_{0.9}$ & (110, 170, 35) & 1104 & 0 & 0 \\
         \bottomrule \\
    \end{tabular}
\end{table}

\begin{figure}
    \centering
    \includegraphics[width=0.86\columnwidth]{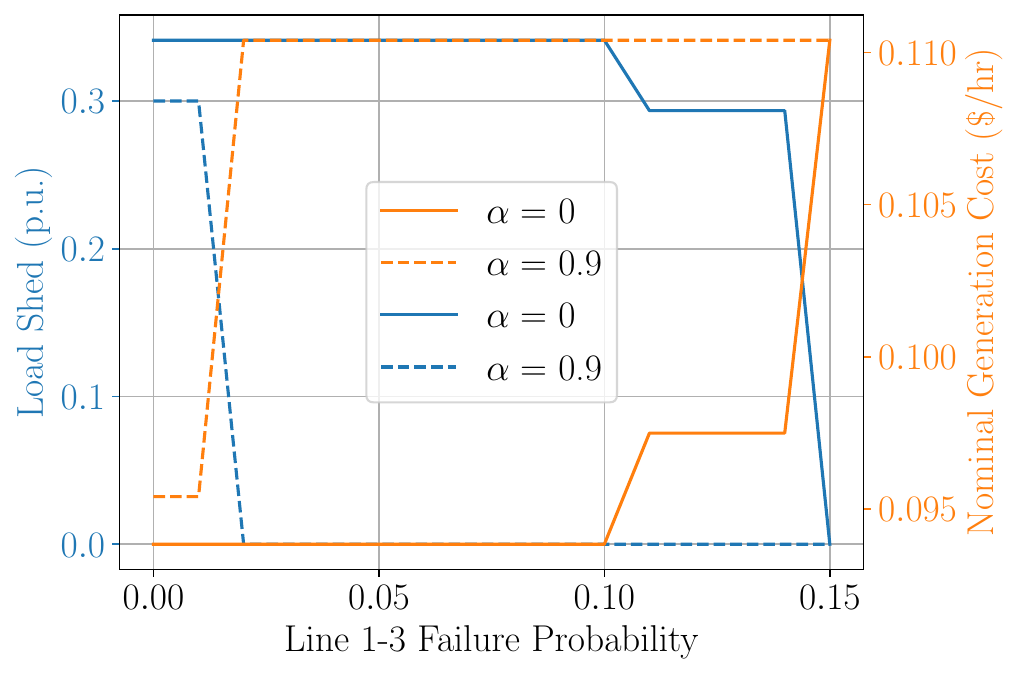}
    \caption{The impact of varying line 1-3 failure probability on nominal generation cost and load shed in the 3-bus example when $\alpha = 0$ (solid) and $\alpha=0.9$ (dotted).}
    \label{fig:3bus_vary_p}
\end{figure}

Notice that when the SO has lower risk aversion, i.e., when $\alpha\in\{0,0.1\}$, she is willing to incur some level of load shed. As she becomes more risk-averse to high costs, expressed as $\alpha=0.9$, the quantity of load shed and reserve capacity decrease and in fact both reach zero, meaning the SO effectively implements \PSCED{}, albeit under relaxed short-term emergency level line flow constraints. The dispatch in this case prioritizes local generation to avoid shortfalls in the event of a contingency.

\revision{Relaxing the assumption that single line failure probabilities are fixed to a common value, Figure \ref{fig:3bus_vary_p} illustrates the impact of varying line failure probability of a single line on the nominal generation cost and amount of load shed.  In particular, we vary the failure probability of line 1-3 in Figure \ref{fig:3bus}, as the other probabilities remain fixed at $p_k=0.1$. The values of each curve plotted do not change for probabilities larger than 0.15, and are thus omitted. 
Higher chances of line failure increase risks of power delivery, compensating for which requires a higher-cost nominal generation. When the SO is highly risk-averse ($\alpha=0.9$), nominal generation costs quickly climb as the failure probability increases from zero, while a risk-neutral SO $(\alpha=0)$  only adjusts nominal generation for large enough failure probability. Moreover, an increase in this probability amplifies the cost associated with load-shed through the $\CVaR{}_{\alpha} \left[ \mathcal{C}(\v{\delta d} ) \right]$ term in \RSCED{}, forcing the algorithm to shed less load. This term increases with the tunable risk parameter $\alpha$. As a result, the drop in the load-shed occurs at smaller values of the line failure probability with $\alpha = 0.9$ compared to that with $\alpha = 0$.
All quantities agree for large enough failure rates ($\geq$ 0.15).}

%% file: 3bus.tex

\begin{figure}[H]
\centering
\begin{tikzpicture}[circuit ee IEC, scale=1]
	\coordinate (b1) at (0, 2);
	\coordinate (b2) at (0, 0);
	\coordinate (b3) at (3, 1);
	\def\busLength{0.5}
	\def\busSep{0.75}
	\def\lineSep{0.4}
	\def\compSep{0.7}
	\def\threeCompBusSep{0.3}
	\def\twoCompBusSep{0.2}
	
	\node [label={above:$1$}] at ($(b1) + (0,\busLength)$) {};
	\node [label={above:$2$}] at ($(b2) + (0,\busLength)$) {};
	\node [label={above:$3$}] at ($(b3) + (0,\busLength)$) {};

	\draw [very thick] ($(b1) + (0,\busLength)$) -- ($(b1) - (0, \busLength)$);
	\draw [very thick] ($(b2) + (0,\busLength)$) -- ($(b2) - (0, \busLength)$);
	\draw [very thick] ($(b3) + (0,\busLength)$) -- ($(b3) - (0, \busLength)$);

	\node [ac source, info=$g_1$, label={below:\small$\$5$/MWh}] at ($(b1) - (\busSep + \busSep, -\twoCompBusSep)$) (g1) {};
	\node [] at ($(b1) - (\busSep, \compSep + \twoCompBusSep)$) (d1) {\small $110$ MW};

	\draw [-] ($(b1) + (0, \twoCompBusSep)$) -- (g1);
	\draw [arrows={-angle 60}] ($(b1) - (0, \twoCompBusSep)$) -| (d1);

	\node [ac source, info=$g_2$, label={below:\small$\$1.2$/MWh}] at ($(b2) - (\busSep + \busSep, -\twoCompBusSep)$) (g2) {};
	\node [] at ($(b2) - (\busSep, \compSep + \twoCompBusSep)$) (d2) {\small $110$ MW};

	\draw [-] ($(b2) + (0, \twoCompBusSep)$) -- (g2);
	\draw [arrows={-angle 60}] ($(b2) - (0, \twoCompBusSep)$) -| (d2);

	\node[ac source, info=$g_3$,, label={below:\small$\$10$/MWh}] at ($(b3) - (-\busSep - \busSep, -\twoCompBusSep)$) (g3) {};
	\node [] at ($(b3) - (-\busSep, \compSep + \twoCompBusSep)$) (d3) {\small $95$ MW};

    \draw [-] ($(b3) + (0, \twoCompBusSep)$) -- (g3);
	\draw [arrows={-angle 60}] ($(b3) - (0, \twoCompBusSep)$) -| (d3);

	\path [draw] ($(b1) + (0, -\twoCompBusSep)$)
				-- ($(b1) + (\lineSep, -\twoCompBusSep)$) 
				-- node [midway,right] {\small$X_{12} = 0.62 p.u.$} ($(b2) + (\lineSep, \twoCompBusSep)$) 
				-- ($(b2) + (0, \twoCompBusSep)$);

	\path [draw] ($(b1) + (0, \twoCompBusSep)$) 
				-- ($(b1) + (\lineSep, \twoCompBusSep)$)
				-- node [midway,above,sloped] {\small$X_{13} = 0.9 p.u.$} ($(b3) - (\lineSep, -\twoCompBusSep)$) 
				-- ($(b3) + (0, \twoCompBusSep)$);

	\path[draw] ($(b2) - (0, \twoCompBusSep)$)
				-- ($(b2) + (\lineSep, -\twoCompBusSep)$)
				-- node [midway,below,sloped] {\small$X_{23} = 0.75 p.u.$} ($(b3) - (\lineSep, \twoCompBusSep)$)
				-- ($(b3) - (0, \twoCompBusSep)$);
\end{tikzpicture}
\caption{A 3-bus network example.}
\label{fig:3bus}
\end{figure}

%% file: pricing.tex
\section{Pricing \RSCED{}-based Dispatch}
\label{sec:pricing}


Pricing \SCED{} problems is a well-known challenge \cite{shi2022scenario}. Desirable properties of LMP-based pricing for the ED problem, such as revenue adequacy for SOs and individual rationality for generators cannot be easily extended to security-constrained counterparts. One way to extend such prices to SCED problems is to establish LMP-style prices based solely upon nominal constraints, an approach known as \LMPnom{}. 
As argued in, e.g., \cite{hogan2013electricity}, 
\LMPnom{} based markets often result in insufficient revenue for appropriate grid maintenance and require large out-of-market settlements to ensure that generators follow the SO's dispatch signals. 

Alternatively, pricing based on the 
\emph{locational marginal price under security} (\LMPmar{})
has been proposed in \cite{galiana2005scheduling}. This model incorporates anticipated congestion across all credible contingencies into nodal prices. In this section, we will study \LMPnom{} and \LMPmar{}, establishing results on revenue adequacy and evaluating the out-of-market settlements required to ensure individual rationality amongst generators. \rev{To the best of our knowledge, such results thus far have been limited to approaches that optimize expected costs of recourse actions, or exclude recourse altogether. We emphasize that the results we present apply to the existing \CSCED{} formulation described in \eqref{eq:csced} as well our \RSCED{} formulation.} 

We use the superscript $\star$ in this section to denote an element of a primal-dual optimal solution of the \RSCED{} problem \eqref{eq:RSCED.lp}. See Appendix \ref{app:proof.thm.ms} for the complete Karush-Kuhn-Tucker (KKT) optimality conditions for problem \eqref{eq:RSCED.lp}. 

\begin{definition}
  The locational marginal prices of the nominal (\LMPnom{}) are defined as
  \begin{equation}
      \v{\pi}^{\N} := \lambda^\star \bone - \v{H}^\top \v{\mu}^\star.
      \label{eq:price.nominal}
  \end{equation}
\end{definition}
In \eqref{eq:price.nominal}, $\lambda^\star$ captures the network energy price, while $\v{H}^{\top}\v{\mu}^{\star}$ captures the congestion prices under nominal conditions. Note that while the form of the LMPs \eqref{eq:price.nominal} is the same as in the standard ED case given in \eqref{eq:ED_LMP}, the Lagrange multipliers arise from different optimization problems, i.e., \eqref{eq:RSCED.lp} and \eqref{eq:ED}, respectively. 
As noted in \cite{hogan2013electricity}, in ignoring congestion costs due to potential contingencies, the \LMPnom{} approach can lead to realized generator dispatches that deviate significantly from their profit-maximization strategies and, as we will see, fail to guarantee revenue adequacy for the SO. 

\begin{definition}
  The locational marginal prices under security (\LMPmar{}) are defined as
  \begin{equation}
    \v{\pi}^{\S} := \lambda^{\star} \bone - \v{H}^\top \v{\mu}^\star - \sum_{k = 1}^K \v{H}_k^\top (\v{\mu}_k^{\DA{}\star} + \v{\mu}_k^{\SE\star}).
    \label{eq:price.marginal}
  \end{equation}
\end{definition}
Unlike the \LMPnom{} pricing, this model explicitly incorporates an additional cost associated with potential recourse actions under the contingency scenarios. \revision{In a sense, \LMPmar{} adds contingency-specific price components to \LMPnom{}.}
The payment in \eqref{eq:price.marginal} is further justified by the fact that akin to the deterministic case, entries of $\v{\pi}^{\S}$
 measure the marginal sensitivity of the optimal cost in \eqref{eq:RSCED.lp} to demand at each node. 

Let $\v{\pi}\in\{\v{\pi}^{\N},\v{\pi}^{\S}\}$. Under either of these pricing schemes, we define the payment made by the demander at bus $i$ as 
$$\Pi^d_i[\v{\pi}] := \pi_i d_i,$$ where $d_i$ is the quantity demanded and $\pi_i$ is the LMP at node $i\in[n]$. Supplier $i\in[n]$ is compensated 
\begin{equation}\label{eq:supp_comp}
    \Pi^g_i[\v{\pi}] = \pi_i g^{\star}_i + \sum_{k=1}^K(\overline{\rho}^{\star}_{k,i}\overline{r}^{\star}_i + \underline{\rho}^{\star}_{k,i} \underline{r}^{\star}_i),
\end{equation} where $g^{\star}_i$ is the quantity produced, $\overline{\rho}^{\star}_{k,i}$ and $\underline{\rho}^{\star}_{k,i}$ are the dual multipliers associated with the reserve capacity requirements in \eqref{eq:RSCED.lp}, and $\ul{r}^{\star}_i$ and $\ol{r}^{\star}_i$ are the reserve capacities procured at node $i\in[n]$. Note that $\overline{\rho}^{\star}_{k,i}$ and $\underline{\rho}^{\star}_{k,i}$ are the marginal costs of reserve capacity at node $i$ in outage scenario $k$. Further, both pricing schemes consist entirely of \emph{ex-ante} payments, i.e., payments are made prior to the realization of a particular scenario and thus do not depend upon the scenario realized. Therefore, the revenue-related results we derive in subsequent sections always apply, rather than in a statistical sense, e.g., in expectation.

\subsection{Revenue Adequacy} 
\label{sub:revenue_adequacy}

We first address the issue of revenue adequacy under the \LMPnom{} and \LMPmar{} pricing schemes. Specifically, we aim to answer whether the merchandising surplus of the SO is nonnegative with these candidate pricing schemes to accompany the dispatch signals computed from the R-SCED problem \eqref{eq:RSCED.lp}. 
Augmenting the expression in \eqref{eq:MS} with additional compensation for generator reserve procurement, the merchandising surplus for pricing scheme $\v{\pi}$ under \RSCED{} is given by
\begin{equation}\label{eq:MS_RSCED}
\begin{split}
    \MS{}[\v{\pi}] &= \sum_{i=1}^n\Pi^d_i[\v{\pi}] - \sum_{i=1}^n\Pi^g_i[\v{\pi}]\\
    &= \v{\pi}^\top \v{d} - \v{\pi}^\top \v{g}^{\star} - \sum_{k=1}^K(\overline{\v{\rho}}_k^{\star\top} \overline{\v{r}}^{\star} + \underline{\v{\rho}}^{\star\top}_k \underline{\v{r}}^{\star}).
\end{split}
\end{equation}

Applying the price definitions \eqref{eq:price.nominal} and \eqref{eq:price.marginal} in \eqref{eq:MS_RSCED}, we have the following result.

\begin{theorem}
  \label{thm:ms}~
          Revenue adequacy is not guaranteed under the \LMPnom{} scheme \eqref{eq:price.nominal}. 
      The \LMPmar{} pricing scheme \eqref{eq:price.marginal}, however, is revenue adequate, i.e., $\MS{}[\v{\pi}^{\S}] \ge 0$.
\end{theorem}
See Section \ref{sec:simulation_results} for a numerical example demonstrating a case wherein $\MS[\pi^{\N}]<0$. In particular, in Figure \ref{fig:24bus.ms.uplift}, $\MS[\pi^{\N}]<0$ for values of $\alpha$ larger than 0.7.
The proof of the claim regarding S-LMP in Theorem \ref{thm:ms} can be found in Appendix \ref{app:proof.thm.ms}. 
Note that the second result in Theorem \ref{thm:ms} holds for all values of the \CVaR{} parameter $\alpha$, i.e., regardless of risk preference, \rev{including the special case where $\alpha = 0$, i.e., when expected load shed is minimized.}
As we will see empirically in Section \ref{sec:simulation_results}, while $\MS{}[\v{\pi}^{\S}]$ is nonnegative, it also generally increases with risk-aversion. 
In the long run, increased merchandising surplus can be used to perform grid upgrades to improve reliability accordingly. \revision{In effect, in ignoring price components from the various contingencies, \LMPnom{} can fail to collect enough rents from demanders to pay the generators--a scenario that can never arise under \LMPmar{}. Reflecting costs from unrealized scenarios makes \LMPmar{} a better pricing scheme than \LMPnom{} based on revenue adequacy properties. However, such a pricing scheme requires a careful selection of modeled contingencies and their probabilities from the SO that market participants must agree upon.}

\subsection{Individual Rationality and Lost Opportunity Cost Payments}\label{sec:IRandLOC}
We now address the question of individual rationality of generators under the \LMPnom{} and \LMPmar{} pricing schemes. That is, are generators being compensated at the same level as their profit-maximizing dispatch, given the price? While implemented via a single stage of ex-ante transactions, the payment schemes that we have presented here have two parts: payment for nominal generation, and compensation for ancillary, scenario-dependent generation by way of payments for reserve procurement. 
As a result, defining what it means to maximize profit is not straightforward. In particular, one must determine whether to account for profit associated with reserve capacity.

The \RSCED{} formulation \eqref{eq:RSCED.lp} and generator compensation structure \eqref{eq:supp_comp}, entails simultaneous settlement of two markets, one for nominal case generation, and another for reserve capacity. Setting aside the concern that the co-existence of these markets may create opportunities for arbitrage or the exercise of market power, as described in e.g., \cite{pritchard2010single} and \cite{borenstein2002measuring}, assessment of the profitability of multi-product firms such as the generators we consider is complicated \cite{hirst2000maximizing,buchsbaum2020spillovers}. 
Depending upon energy and reserve capacity prices, as well as equipment characteristics and operating costs, generators may be incentivized to supply more heavily in one market over another, potentially producing inefficient outcomes.

Instead, we assume that generators act as \emph{energy-only} profit-maximizers solely in the nominal case, viewing reserve procurement as an additional ``bonus'' sum. In this way, we can restrict attention to settlements for nominal generation. 
For the nominal case, if generators are asked to act contrary to their energy-only profit-maximizing dispatch, given the price, they must be paid a \emph{lost opportunity cost} (\LOC{}) payment corresponding to lost revenue due to deviation from their individually optimal dispatch, described below.

The \LOC{} payment for a generator at bus $i$ is calculated as the difference in the energy-only profit from the prescribed dispatch, $(\pi_i - c_i) g_i^\star$, and the energy-only profit-maximizing dispatch with prices derived from the solution of \eqref{eq:RSCED.lp}, given by
\begin{equation}
  \label{eq:ind.rat}
\begin{aligned}
    & \underset{\hat{g}_i}{\text{maximize}} && (\pi_i - c_i) \hat{g}_i, & \text{subject to} && \underline{g}_i \le \hat{g}_i \le \overline{g}_i.
\end{aligned}
\end{equation}
The optimal solution to \eqref{eq:ind.rat}, given price $\pi_i$ is 
  \begin{equation}\label{eq:opt_gen_sol}
      \hat{g}_i^\star := \begin{cases}
        \overline{g}_i & \pi_i \geq c_i, \\
        \underline{g}_i & \pi_i < c_i. 
      \end{cases}
  \end{equation}
Thus, defining the difference between dispatched generation and \eqref{eq:opt_gen_sol} as
\begin{equation}
      \Gamma_i[\v{\pi}] := \begin{cases}
        \overline{g}_i - g_i^\star & \pi_i \ge c_i, \\
        g_i^\star - \underline{g}_i & \pi_i < c_i, 
      \end{cases}
  \end{equation}
we have that $\LOC{}_i[\v{\pi}]$, the \LOC{} payment to generator $i$ under pricing scheme $\pi$ is given by 
\begin{equation}\label{eq:LOC_i}
      \LOC{}_i[\v{\pi}] = | \pi_i - c_i | \Gamma_i[\v{\pi}]. 
  \end{equation}


Summing \eqref{eq:LOC_i} over $i$ then gives the total \LOC{} payments SOs must pay under a given pricing scheme. Using the KKT condition in \eqref{eq:rsced.kkt.stat.gr}, and price definitions \eqref{eq:price.nominal} and \eqref{eq:price.marginal}, we have 
{\small
\begin{equation}\label{eq:loc_pi_N}
    \begin{split}
        &\LOC{}[\v{\pi}^{\N}]\\
        &= \left| \overline{\v{\gamma}}^{\star} - \underline{\v{\gamma}}^{\star} + \sum_{k = 1}^K \v{H}_k^\top (\v{\mu}_k^{\DA{}\star} + \v{\mu}_k^{\SE{}\star}) + \overline{\v{\gamma}}^{\star}_k - \underline{\v{\gamma}}^{\star}_k \right|^\top \v{\Gamma}[\v{\pi}^{\N}],
    \end{split}
\end{equation}
\begin{equation}\label{eq:loc_pi_S}
    \LOC{}[\v{\pi}^{\S}]=\left| \overline{\v{\gamma}}^{\star} - \underline{\v{\gamma}}^{\star} + \sum_{k = 1}^K \overline{\v{\gamma}}^{\star}_k - \underline{\v{\gamma}}^{\star}_k \right|^\top \v{\Gamma}[\v{\pi}^{\S}].
\end{equation}
}
Notice that the $\LOC[\v{\pi}^{\N}]$ expression in \eqref{eq:loc_pi_N} includes additional scenario-dependent congestion components $\v{H}_k^\top (\v{\mu}_k^{\DA{}\star} + \v{\mu}_k^{\SE{}\star})$ for each $k$. Intuitively speaking, as an SO becomes increasingly risk-averse, optimal values of the \RSCED{} objective in \eqref{eq:RSCED.lp} become more sensitive to worst-case scenarios. This sensitivity is reflected in the magnitude of these scenario-dependent congestion components. 
As a result, $\v{\mu}_k^{\DA{}\star}$ and $\v{\mu}_k^{\SE{}\star}$ are expected to grow significantly, resulting in increased $\LOC{}[\v{\pi}^{\N}]$. 

The value of these \LOC{} payments is highly relevant to the previous study of the \MS{}. SOs seek pricing schemes for which the \emph{total revenue}, i.e., \MS{} less aggregate \LOC{} payments, is non-negative. To that end, we  provide the following result for \LMPmar{}, the proof of which is included in Appendix \ref{app:proof.thm.revenue}.
\begin{theorem}
  \label{thm:revenue}
  The total revenue of a SO under the pricing scheme \LMPmar{} is positive, or
  \begin{equation}
      \MS{}[\v{\pi}^{\S}] - \bone^\top \LOC{}[\v{\pi}^{\S}] \ge 0.
  \end{equation}
\end{theorem}
Thus, even while ensuring individual rationality via \LOC{} payments, SOs are able to operate with the pricing scheme \LMPmar{} without fear of operating at a deficit. Intuitively speaking, under the \LMPmar{} scheme, payments from demanders and to suppliers account for contingencies, i.e., the market clearing process itself sufficiently accounts for reserve procurement and recourse actions. By Theorem \ref{thm:ms}, this process yields a nonnegative merchandising surplus, which the proof of Theorem \ref{thm:revenue} shows is ample to cover the remaining \LOC{} payments. The \LMPnom{} scheme, on the other hand, does not necessarily yield nonnegative \MS{}. Furthermore, \LMPnom{} \LOC{} payments have a congestion component that is expected to increase with risk-aversion, but no corresponding increase in revenue. This is problematic as out-of-market settlements may exceed the merchandising surplus. In fact, we will observe this phenomenon in our empirical analysis of the IEEE 24-bus RTS network in Section \ref{sec:simulation_results}. \revision{Again, similar to our observation for merchandising surplus, \LMPmar{} fares better than \LMPnom{} in ensuring that rents collected from demanders can sufficiently incentivize the generators to follow the SO-intended dispatch. Allowing prices to reflect what might occur in contingency scenarios helps to pay generators to follow a nominal dispatch whose calculation through R-SCED already captures the effect of said contingencies. In a way, \LMPmar{} ties the pricing mechanism more closely to the process through which the dispatch is computed than does \LMPnom{}.}

%% file: algorithm.tex
\begin{algorithm}[t]
	\setlength{\belowdisplayskip}{0pt} \setlength{\belowdisplayshortskip}{0pt}
	\setlength{\abovedisplayskip}{0pt} \setlength{\abovedisplayshortskip}{0pt}
	\kwInit{Choose $\v{x}_0^1 \in \Xset$, $t_k^\star = -\infty,  k\in[K]$.}
	\For{$i = 1, 2, \dots$}{ \For{$k \in[K]$}{Set $\v{\lambda}^{i}_k$, $J_k^i$ as the optimizer, optimum of\begin{equation}
		\begin{aligned}& \underset{\v{\lambda}_k\geq 0}{\text{maximize}} && -\v{b}_k^T \v{\lambda}_k + \v{\lambda}_k^T \v{A}_k \v{x}_0^i,
			\\
			& \text{subject to} && \v{c}_k + \v{E}_k^T \v{\lambda}_k = 0.
		\end{aligned}
		\end{equation}
		\label{alg:bender.1}
    }
		\label{alg:bender.term}
		\If{$J_k^i = t_k^\star \; \forall k\in[K]$}{
        Terminate.}
        Set $(\v{x}_0^{i + 1}, t_1^\star, \dots, t_k^\star)$ as the optimizers of
		\begin{equation}\nonumber 
		\begin{aligned}
			& \underset{\substack{\v{x}_0 \in \Xset, \\ t_1, \dots, t_K}}{\text{minimize}} && \v{c}^T \v{x}_0 + \frac{1}{1 - \alpha} \sum_{k = 1}^K t_k, \\
			& \text{subject to} && t_k \ge J_k^{\ell} + (\v{x}_0 - \v{x}_0^{\ell})^{\top}\v{A}^{\top}_k\v{\lambda}_k^{\ell},
			\\ &&&
			 \ell \in [i], \; k \in [K].
		\end{aligned}
		\end{equation}
		\label{alg:bender.3}
	}
\caption{Benders' decomposition for \eqref{eq:RSCED.decomposable}.}
\label{alg:bender}
\end{algorithm}

\section{Solution via Benders' Decomposition}
\label{sec:algorithm}

As detailed in Section \ref{sec:formulating}, \RSCED{} can be written as a large linear program. Collecting the decision variables into vectors for all $k\in [K]$ with
%
\begin{equation}\nonumber
    \v{x}_0 := \begin{pmatrix} z & \v{g}^{\top} & \underline{\v{r}}^{\top} & \overline{\v{r}}^{\top} \end{pmatrix}^{\top}, \quad \v{x}_k := \begin{pmatrix} y_k & \v{\delta g}^{\top}_k & \v{\delta d}^{\top}_k \end{pmatrix}^{\top},
\end{equation}
%
 problem \eqref{eq:RSCED.lp} can now be written as
\begin{equation}
\begin{aligned}
    &\underset{\v{x}_0, \v{x}_1, \dots, \v{x}_k}{\text{minimize}} && \v{c}^\top \v{x}_0 + \frac{1}{1 - \alpha} \sum_{k = 1}^K \v{c}_k^\top \v{x}_k, \\
    &\text{subject to} && \v{A} \v{x}_0 \le \v{b}, \\
    &&& \v{A}_k \v{x}_0 + \v{E}_k \v{x}_k \le \v{b}_k, \; k\in[K],
\end{aligned}
\label{eq:RSCED.decomposable}
\end{equation}
for appropriate $\v{c}$, $\v{c}_k$, $\v{A}$, $\v{b}$, $\v{A}_k$, $\v{E}_k$, and $\v{b}_k$.
While linear programs such as \eqref{eq:RSCED.decomposable} admit polynomial-time algorithms, the large scale of these problems for practical power systems results in very high dimensionality, demonstrated in Table \ref{tab:ED.dim}. This results in significant time complexity, rendering the problems intractable within the time periods typically required for making dispatch decisions. \revision{Fortunately, the formulation in \eqref{eq:RSCED.decomposable} is amenable to decomposition whose structure can be exploited to parallelize and speed up computation.}

\begin{table}[ht]
\centering
	\caption{Problem dimensions for the IEEE 200-bus network.}
	\begin{tabular}{l c c}
		\toprule
		\textbf{Formulation} & \textbf{Variables} & \textbf{Constraints} \\ 
		\midrule 
		ED & 49 & 589 \\
		\PSCED & 49 & 120,639 \\
		\CSCED & 61,054 & 386,954 \\
		\RSCED & 61,398 & 411,454 \\
		\bottomrule \\
	\end{tabular}
	\label{tab:ED.dim}
\end{table}
%
\revision{With $\Xset := \{ \v{x}_0 | \v{A} \v{x}_0 \leq \v{b} \}$,
\eqref{eq:RSCED.decomposable} can be written as
\begin{equation}
\begin{aligned}
    &\underset{\substack{\v{x}_0 \in \Xset, \\ {t}_1, \dots, {t}_K}}{\text{minimize}} && \v{c}^\top \v{x}_0 + \frac{1}{1 - \alpha} \sum_{k = 1}^K t_k, \\
    &\text{subject to} && t_k \geq J_k^\star(\v{x}_0), k \in [K],
\end{aligned}
\label{eq:decomp.sup.epi}
\end{equation}
}
where
\begin{equation}
\begin{aligned}
    J_k^\star(\v{x}_0) :=\ &\underset{\v{x}_k}{\text{minimum}} && \v{c}_k^\top \v{x}_k, \\
    &\text{subject to} && \v{A}_k \v{x}_0 + \v{E}_k \v{x}_k \le \v{b}_k.
\end{aligned}
\label{eq:decomp.sub}
\end{equation}
\revision{Here, \eqref{eq:decomp.sup.epi} defines the primary problem, while \eqref{eq:decomp.sub} defines the sub-problems that can be solved in parallel for a given value of $\v{x}_0$ from the primary problem. Algorithms such as Benders' decomposition in \cite{benders1962partitioning} and critical region exploration in \cite{guo2017robust}  take advantage of this decomposition. We focus on the former and present it formally in Algorithm \ref{alg:bender}. For completeness, we describe the mechanics behind the algorithm. See \cite{geoffrion1972generalized} for a more detailed treatment.}

\revision{Solving the primary problem requires knowledge of $J_k^\star$ for all $k \in [K]$. Instead, the Benders' decomposition algorithm constructs sequentially tighter lower bounds for $J_k^\star(\v{x}_0)$ via \emph{dual cuts} and solves the primary problem repeatedly with these lower bounds. To explain the process, associate a Lagrange multiplier $\v{\lambda}_k$ to the constraint in \eqref{eq:decomp.sub} and use strong duality of linear programming 
to write
}
\begin{equation}
\begin{aligned}
    J_k^\star(\bar{\v{x}}_0) = \; &\underset{\v{\lambda}_k\geq 0}{\text{maximum}} && -\v{b}_k^\top \v{\lambda}_k + \bar{\v{x}}_0^\top \v{A}_k^\top \v{\lambda}_k, \\
    &\text{subject to} && \v{c}_k + \v{E}_k^\top \v{\lambda}_k = 0
\end{aligned}
\label{eq:decomp.sub.dual}
\end{equation}
for an arbitrary $\bar{\v{x}}_0 \in \Xset$. 
\revision{The optimization problem in the right-hand side of \eqref{eq:decomp.sub.dual} is the dual problem of the sub-problem in \eqref{eq:decomp.sub}.
If $\v{\lambda}_k^\star$ is an optimizer of this dual problem, then
\begin{align}
    J_k^\star(\bar{\v{x}}_0) = -\v{b}_k^\top \v{\lambda}_k^\star + \bar{\v{x}}_0^\top \v{A}_k^\top \v{\lambda}_k^\star.
    \label{eq:benders.dual.opt}
\end{align}
For any $\v{x}_0 \in \Xset$ that is not necessarily the same as $\bar{\v{x}}_0$,
\begin{equation}
\begin{aligned}
    J_k^\star(\v{x}_0)
    &\ge -\v{b}_k^\top \v{\lambda}_k^\star + \v{x}_0^\top \v{A}_k \v{\lambda}_k^\star 
    \\
    &= J_k^\star(\bar{\v{x}}_0) + (\v{x}_0 - \bar{\v{x}}_0)^\top \v{A}_k^\top \v{\lambda}_k^\star,
\end{aligned}
\label{eq:benders.hyperplane}
\end{equation}
where the first step follows from the feasibility of $\v{\lambda}_k^\star$ in the dual problem \eqref{eq:decomp.sub.dual} to compute  $J_k^\star(\v{x}_0)$ and the second step uses \eqref{eq:benders.dual.opt}. Thus, $J_k^\star(\v{x}_0)$ lies above all possible ``cuts'' defined in \eqref{eq:benders.hyperplane} for any $\bar{\v{x}}_0 \in \Xset$.
Benders' decomposition advocates replacing the constraint $t_k \geq J_k^\star({\v{x}}_0)$ in the primary problem \eqref{eq:decomp.sup.epi} with a growing collection of cuts. At each iteration, a new cut is generated using the right-hand side of \eqref{eq:benders.hyperplane} for each $k\in [K]$ with $\bar{\v{x}}_0$ being an optimizer of the same problem in the previous iteration.}



\revision{As shown in \cite{benders1962partitioning}, Benders' decomposition converges to an optimal solution of \eqref{eq:RSCED.decomposable}, provided that each subproblem \eqref{eq:decomp.sub} has a feasible solution for all $\v{x}_0 \in \Xset$.} 
Such an assumption is difficult to guarantee and, in fact, is not typically satisfied for the \RSCED{} formulation in \eqref{eq:RSCED.lp}. One can introduce feasibility cuts to tackle infeasible sub-problems as in \cite{grothey1999note}. 
In our simulations, we add heavily-penalized slack variables to the inequality constraints in \eqref{eq:RSCED.lp} to tackle infeasibility.

Benders' decomposition has been previously applied to \SCED{} problems in \cite{liu2015computational,shi2022scenario}. While our \CVaR-sensitive $\RSCED$ formulation generalizes \CSCED{} in the sense that load-shed may be allowed in some scenarios depending upon SO's risk aversion, \revision{this generalization does not fundamentally complicate algorithm design.}


%% file: simulation.tex
\section{Simulation Results} 
\label{sec:simulation_results}

\begin{figure*}[!ht]
    \centering
    \subfloat[\label{fig:24bus.cost.load}]{\includegraphics[height=1.5in]{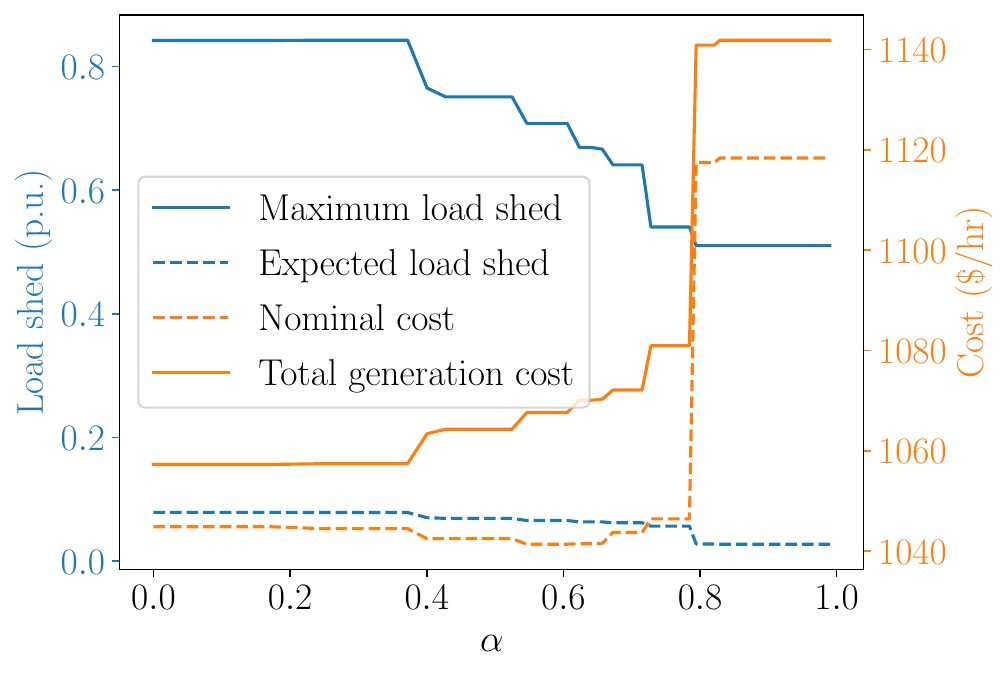}}
    \subfloat[\label{fig:24bus.ms.uplift}]{\includegraphics[height=1.5in]{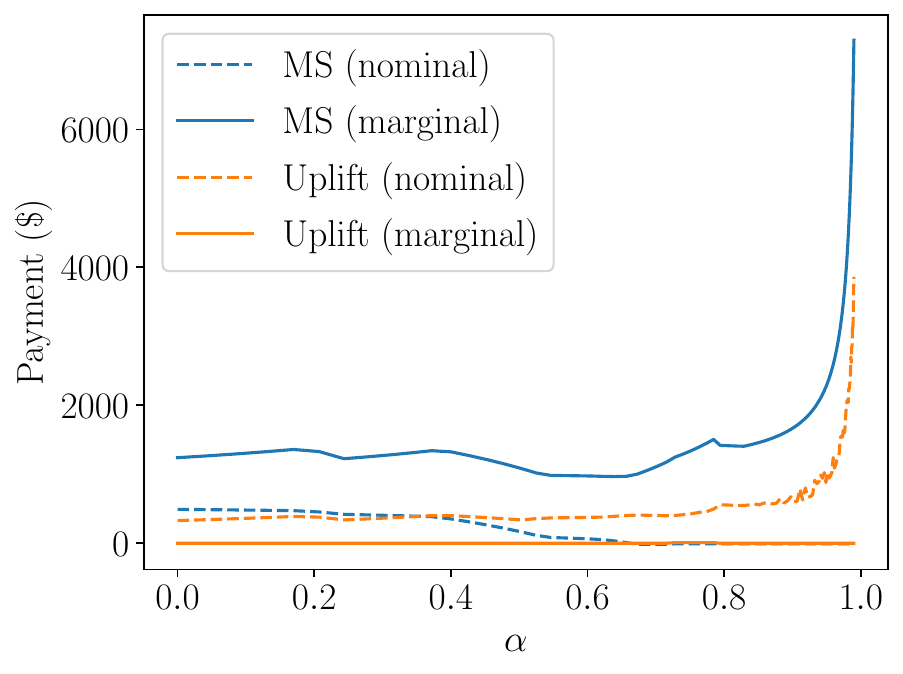}}
    \subfloat[\label{fig:24bus.lmp.alpha06}]{\includegraphics[height=1.6in]{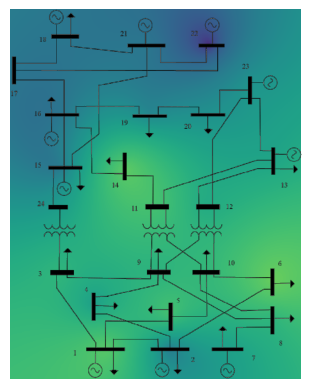}}
    \subfloat[\label{fig:24bus.lmp.alpha09}]{\includegraphics[height=1.6in]{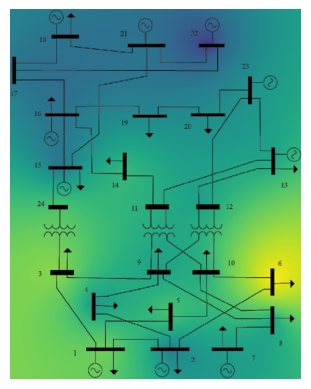}}
    \includegraphics[height=1.6in]{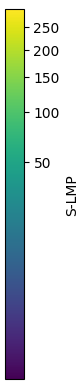}
    \caption{Simulation results on the IEEE 24-bus RTS test network with \protect\subref{fig:24bus.cost.load} load shed (average and total) and dispatch cost (nominal and total including reserve), \protect\subref{fig:24bus.ms.uplift} merchandising surplus and uplift payments under \LMPnom{} and \LMPmar{} schemes, and \LMPmar{} for \protect\subref{fig:24bus.lmp.alpha06} $\alpha = 0.6$, and \protect\subref{fig:24bus.lmp.alpha09} $\alpha = 0.9$.}
    \label{fig:24bus.sim}
    \label{fig:24bus.network}
\end{figure*}



We now present results from the numerical experiments of our \RSCED{} formulation on the IEEE 24-bus RTS test network shown in Figure \ref{fig:24bus.network} from the PGLib v21 \cite{pglib2021}. The network is augmented with a uniform \VoLL{} of $\$9100/$MW, and relaxed line ratings of $\v{f}^{\DA} = 1.7 \v{f}$ and $\v{f}^{\SE} = 1.2 \v{f}$. \rev{Our scenario set consists of every single-line failure, and we assume a uniform distribution over these scenarios.} As seen in Figure \ref{fig:24bus.cost.load}, increasing risk-aversion results in a decrease in the expected and total load shed. In this example, there does not exist a solution without load shed and thus, the canonical \CSCED{} formulation in \eqref{eq:csced} that does not model load shed, would be infeasible. However, the \RSCED{} problem is able to find a solution. As Figure \ref{fig:24bus.cost.load} illustrates, while increasing risk-aversion, i.e., larger $\alpha$ values, corresponds to decreased load shed, it also results in greater dispatch cost, including nominal and reserve capacities. 

Under \LMPnom{}, as illustrated in Figure \ref{fig:24bus.ms.uplift}, the merchandising surplus is not positive for $\alpha>0.7$, an instance of the potential lack of revenue adequacy described in Theorem \ref{thm:ms} (i). As discussed in Section \ref{sec:IRandLOC}, \LMPnom{} LOC payments grow rapidly for $\alpha>0.8$ as the SO becomes more averse to worst-case scenario congestion. For $\alpha>0.4$, merchandising surplus is less than the required LOC payments, meaning that the SO runs cash negative after settling the payments of the market participants. On the other hand, the corresponding \LMPmar{} merchandising surplus is always positive and greater than the LOC payments, affirming the result in Theorem \ref{thm:revenue}. 
In fact, the LOC payments under \LMPmar{} are identically zero for this example across all values of $\alpha$. While we cannot guarantee that this is true in general, the payments are substantially lower using \LMPmar{} than \LMPnom{}, and remain similar across all levels of risk aversion.

\revision{In Figures \ref{fig:24bus.lmp.alpha06} and \ref{fig:24bus.lmp.alpha09}, we provide two heat-maps of \LMPmar{}s across the 24-bus system for two different values of the risk parameter $\alpha$. 
Notice that the prices at certain buses (e.g., at bus 6 in the lower righthand corner of the network) are more sensitive to increases in the tunable parameter $\alpha$ than others (e.g., bus 19 in the upper center part of the network). 
Under the classical LMP pricing mechanism (defined as $\v{\pi}^{\ED}$ in \eqref{eq:ED_LMP}) variation in prices across network buses follows from line congestion at the optimal dispatch, which makes it cheaper to service incremental demand at certain locations than others. 
When that dispatch is calculated using the \RSCED{} formulation, \LMPmar{}-based prices are further affected by \emph{possible} congestion that might arise under optimal recourse decisions in contingency scenarios. Since \RSCED{} balances nominal dispatch, recourse costs, and their probabilities in a unified framework, network price variations, in a sense, reflect the combined effect of network constraints and responses to uncertain contingencies. When SO's risk tolerance is low (with higher $\alpha$), meeting incremental load at one location can be much more costly than at another, as both the nominal dispatch may vary substantially to meet the stringent risk requirements, as well as the congestion patterns in those scenarios. Just as congestion components of LMPs in $\v{\pi}^{\ED}$ can be used as signals for identifying candidate network sections for transmission expansion, maintenance, and upgrades, the congestion components of \RSCED{} \emph{across} scenarios can serve the same purpose. In addition, \LMPmar{}-based prices may also highlight the line failure scenarios that contribute more heavily. }

\begin{table}[h]
    \centering
    \caption{The effect of adjusting the line failure probability of the line between buses 3 and 24 with $\alpha = 0.6$.}
    \label{tab:24bus.prob.compare}
    \begin{tabular}{l c c}
         \toprule
         \textbf{Failure} & \textbf{Nominal}
         & \textbf{Max load} \\
         \textbf{Probability} & \textbf{cost (\$/h)}
         & \textbf{shed (p.u.)}  \\
         \midrule
         $0.$ & 1042.854 & 1.7026 \\
         $0.001$ & 1042.941 & 0.6994 \\
         $0.01$ & 1113.926 & 0.5104 \\
         \bottomrule \\
    \end{tabular}
\end{table}

\revision{In Table \ref{tab:24bus.prob.compare}, we study the impact of changing the line failure probability of a single line (joining buses 3 and 24), while holding the probabilities of other line failure scenarios to 0.0014. The results display the same trends as observed for the 3-bus system in Section \ref{sec:rsced}.}

\subsection{Evaluating Benders' decomposition for arbitrary networks}

To demonstrate the viability of the Benders' decomposition approach to solving the \RSCED{} problem for practical networks, we evaluate the solution time for a number of PGLib v21 networks. \rev{As in the previous simulation, we take as our scenario set every possible single-line failure, and assume a uniform distribution over these scenarios}. The run-times are shown in Table \ref{tab:runtimes}, where the column labeled ``Gurobi" lists runtimes for solving the large linear program in \eqref{eq:RSCED.lp} using Gurobi. As the table indicates, run-times for both the large LP solution and Benders' decomposition generally increase with network size. Benders' decomposition becomes competitive with solving the large LP for systems larger than 14 buses and significantly 
faster for networks larger than 73 buses.
\begin{table}[ht]
    \centering
    \caption{Runtime comparison of Benders' Decomposition and direct solution of the large LP formulation of the \RSCED{} problem.}
    \begin{filecontents*}{data/rsced_runtimes.csv}
casename,g0,b0,g9,b9
case3_lmbd,0.0016,0.0092,0.0006,0.0010
case3_lmbd__api,0.0007,0.0008,0.0005,0.0016
case3_lmbd__sad,0.0009,0.0013,0.0007,0.0010
case5_pjm,0.0023,0.0024,0.0021,0.0024
case5_pjm__api,0.0021,0.0022,0.0022,0.0029
case5_pjm__sad,0.0022,0.0034,0.0023,0.0020
case14_ieee,0.0078,0.0122,0.0078,0.0083
case14_ieee__api,0.0080,0.0091,0.0085,0.0087
case14_ieee__sad,0.0073,0.0082,0.0073,0.0081
case24_ieee_rts,0.2078,0.0473,0.1935,0.0445
case24_ieee_rts__api,0.2527,0.0990,0.2445,0.0986
case24_ieee_rts__sad,0.1931,0.0456,0.1909,0.0484
case30_as,0.0755,0.0407,0.0742,0.0422
case30_ieee,0.0465,0.0381,0.0526,0.0393
case30_as__api,0.0913,0.0641,0.0894,0.7291
case30_ieee__api,0.0489,0.0442,0.0487,0.0443
case30_as__sad,0.0697,0.0387,0.0687,0.0358
case30_ieee__sad,0.0407,0.0399,0.0419,0.0400
case39_epri,0.1436,0.0658,0.1341,0.0611
case39_epri__api,0.1586,0.0806,0.1586,0.0870
case39_epri__sad,0.1263,0.0607,0.1266,0.0619
case57_ieee,0.3670,0.2721,0.3643,0.2631
case57_ieee__api,0.3904,0.2048,0.3827,0.2124
case57_ieee__sad,0.3700,0.2610,0.3753,0.2609
case73_ieee_rts,6.3827,1.8741,7.2492,1.2606
case73_ieee_rts__api,10.4875,2.0926,13.5974,2.4620
case73_ieee_rts__sad,6.8249,1.2365,6.5114,1.2421
case118_ieee,17.0832,4.7362,15.7959,4.8882
case118_ieee__api,17.0718,5.0415,17.2200,4.9935
case118_ieee__sad,14.7307,3.7659,14.7360,3.7275
case200_activ,17.4605,4.7793,18.4642,4.5208
case200_activ__api,28.8419,9.7479,28.5164,9.3644
case200_activ__sad,17.1610,4.7703,16.7447,4.3801
\end{filecontents*}
\csvreader[column names={casename=\cname,g0=\gzero,b0=\bzero,g9=\gnine,b9=\bnine},tabular=lrrrr,head,respect underscore=true,table head=\toprule & \multicolumn{2}{c}{\uline{\hspace{0.6cm}${\alpha = 0}$\hspace{0.6cm}}} & \multicolumn{2}{c}{\uline{\hspace{0.5cm}$\alpha = 0.9$\hspace{0.5cm}}} \\ Case Name & Gurobi & Benders' & Gurobi & Benders' \\ \midrule, table foot=\bottomrule \\]{data/rsced_runtimes.csv}{}{\cname & \gzero & \bzero & \gnine & \bnine }
    \label{tab:runtimes}
    \setlength\belowcaptionskip{-1in}
\end{table}
%

\rev{The results presented in Table \ref{tab:runtimes} constitute a proof of concept, demonstrating that our formulation is amenable to decomposition schemes such as Benders'. 
Algorithmic enhancements to the base Benders' approach (e.g., in \cite{liu2015computational}) will likely improve run-times over the results in Table \ref{tab:runtimes}, and facilitate solution of \RSCED{} over larger power networks.} \revision{We emphasize that \RSCED{}, with its flexibility in specifying SO's risk preference, falls in the same optimization class as \CSCED{}. In addition, \RSCED{} admits the decomposition in \eqref{eq:decomp.sup.epi}-\eqref{eq:decomp.sub} that shares parallels with that for \CSCED{} in \cite{liu2015computational}, permitting algorithmic innovations (including and beyond Benders') to apply seamlessly across formulations.}


%% file: conclusion.tex


\section{Conclusion} 
\label{sec:conclusion}

In this paper, we studied a risk-sensitive security-constrained economic dispatch formulation that allows an SO to trade off between the nominal dispatch cost and reliability of meeting demand. \rev{In this risk-sensitive context}, we studied two pricing mechanisms, one based solely on the nominal dispatch and another based on the total marginal cost of demand. We established results on revenue adequacy, and evaluated the lost opportunity cost payments under either scheme, \rev{closing a gap in understanding of market properties of these mechanisms under risk aversion. Our results on pricing policies apply to existing \CSCED{} formulations as well. Despite offering a generalization over prior approaches in the inclusion of an SO's risk attitudes, our formulation remains amenable to solution via decomposition techniques such as Benders'.} We studied the effect of risk aversion on the \RSCED{} dispatch and the pricing schemes in an IEEE 24-bus RTS network, and demonstrated the utility of Benders' decomposition in reducing solution time of \RSCED{} over large power networks.

\revision{While we concretely presented a framework to optimize energy and reserve procurement under single line failure scenarios, our risk-sensitive design framework extends to any forward market context that must factor in possible contingencies, their probabilities, the available recourse actions in said contingencies, and their associated costs. Such a framework will allow the system to successfully respond to scenarios that might arise during real-time operations.
Our analysis revealed that a pricing mechanism that included components due to costs associated with recourse decisions in such contingencies fared better than another that ignored them. These components help the market settlements better reflect the value of the response to the uncertain scenarios. As the power system becomes increasingly complex, the number of possible scenarios will undoubtedly increase, growing the computational burden on the market clearing mechanism. A judicious scenario screening procedure will then play a central role in the efficacy of the overall market design. Studying the result of explicit incorporation of unit commitment decisions and scenario selection procedures in \RSCED{} remains an interesting direction for future endeavors.}


%% file: proofs.tex

\input{KKT}

\subsection{Proof of Theorem \ref{thm:ms}}
\label{app:proof.thm.ms}
Applying definition \eqref{eq:price.marginal} to \eqref{eq:MS_RSCED}, we have
{\small
\begin{flalign*}
  &\MS{}[\v{\pi}^{\S}] = \left[\lambda^{\star} \bone^\top - \v{\mu}^{\star\top} \v{H} - \sum_{k = 1}^K (\v{\mu}_k^{\DA{}\star} + \v{\mu}_k^{\SE{}\star})^\top \v{H}_k\right] (\v{d} - \v{g}^{\star}) \\
    &\quad - \sum_{k = 1}^K [\underline{\v{\rho}}_k^{\star\top} \underline{\v{r}}^{\star} + \overline{\v{\rho}}_k^{\star\top }\overline{\v{r}}^{\star}] 
  \end{flalign*}
  }
 {\small
\begin{flalign*} 
	&=\v{\mu}^{\star\top} \v{f} + \sum_{k = 1}^K (\v{\mu}_k^{\DA{}\star} + \v{\mu}_k^{\SE{}\star})^\top \v{H}_k (\v{g}^{\star} - \v{d}) 
 + [\underline{\v{\rho}}^{\star}_k - \overline{\v{\rho}}^{\star}_k]^\top \v{\delta g}^{\star}_k \\
	&= \v{\mu}^{\star\top} \v{f} + \sum_{k = 1}^K [\v{\mu}_k^{\DA{}\star}]^\top \v{f}_k^{\DA{}} + [\v{\mu}_k^{\SE{}\star}]^\top \v{f}_k^{\SE{}} \\
	&\quad + \sum_{k = 1}^K - [\v{\mu}_k^{\SE{}\star}]^\top \v{H}_k (\v{\delta g}^{\star}_k + \v{\delta d}^{\star}_k) + [\underline{\v{\rho}}^{\star}_k - \overline{\v{\rho}}^{\star}_k]^\top \v{\delta g}^{\star}_k,
\end{flalign*}
}
where we use \eqref{eq:RSCED.nominal}, \eqref{eq:rsced.kkt.cs.g}, \eqref{eq:rsced.kkt.cs.flow}, \eqref{eq:rsced.kkt.cs.dg}. By the dual feasibility condition, and positivity of $\v{f}, \v{f}_k^\DA{}$, $\v{f}_k^\SE{}$, we have
{\small \begin{equation*}
    \MS{}[\v{\pi}^{\S}] \ge \sum_{k = 1}^K - [\v{\mu}_k^{\SE{}\star}]^\top \v{H}_k (\v{\delta g}^{\star}_k + \v{\delta d}^{\star}_k) + [\underline{\v{\rho}}^{\star}_k - \overline{\v{\rho}}^{\star}_k]^\top \v{\delta g}^{\star}_k.
    \label{eq:ms.bound.1}
\end{equation*}} 
Applying \eqref{eq:rsced.kkt.stat.d} and \eqref{eq:RSCED.se.balance}, this yields
{\small \begin{equation*}
\begin{aligned}
  &\MS{}[\v{\pi}^{\S}] \ge \sum_{k = 1}^K -\lambda^{\star}_k \bone^\top \v{\delta g}^{\star}_k - \lambda^{\star}_k \bone^\top \v{\delta d}^{\star}_k + (\overline{\v{\gamma}}^{\star}_k - \underline{\v{\gamma}}^{\star}_k)^\top \v{\delta g}^{\star}_k 
  \end{aligned}
  \end{equation*}}
\begin{equation*}
\begin{aligned}
  &\quad + \sum_{k = 1}^K (\overline{\v{\sigma}}^{\star}_k - \underline{\v{\sigma}}^{\star}_k)^\top \v{\delta d}^{\star}_k + \overline{\nu}^{\star}_k \voll^\top \v{\delta d}^{\star}_k \\
  &= \sum_{k = 1}^K (\overline{\v{\gamma}}^{\star}_k - \underline{\v{\gamma}}^{\star}_k)^\top \v{\delta g}^{\star}_k + (\overline{\v{\sigma}}^{\star}_k - \underline{\v{\sigma}}^{\star}_k)^\top \v{\delta d}^{\star}_k + \overline{\nu}^{\star}_k \voll^\top \v{\delta d}^{\star}_k.
  \label{eq:ms.bound.2}
\end{aligned}
\end{equation*}
Using \eqref{eq:rsced.kkt.cs.dg} and \eqref{eq:rsced.kkt.cs.rdcvar}, we have
{\small \begin{equation}
\begin{aligned}
  &\MS{}[\v{\pi}^{\S}] \ge \sum_{k = 1}^K \overline{\v{\gamma}}_k^{\star\top} [ \overline{\v{g}} - \v{g}^{\star} ] + \underline{\v{\gamma}}_k^{\star\top} [ \v{g}^{\star} - \underline{\v{g}} ] + \overline{\v{\sigma}}^{\star}_k \v{\Delta}_d + \overline{\nu}^{\star}_k \voll^\top \v{\delta d}^{\star}_k.
  \label{eq:ms.bound.3}
\end{aligned}
\end{equation}}
The rest follows from the primal feasibility condition in \eqref{eq:RSCED.nominal}, the dual feasibility conditions and positivity of $\v{\Delta}_d, \voll, \v{\delta d}_k$.

\subsection{Proof of Theorem \ref{thm:revenue}}
\label{app:proof.thm.revenue}
For notational convenience, we drop the superscript in $\v{\pi}^{\S}$. From \eqref{eq:LOC_i}, we have
{\small \begin{equation}
\begin{aligned}
  \LOC_i[\v{\pi}] = \max\{
  (\pi_i - c_i) (\overline{g}_i - g_i^\star), 
  (\pi_i - c_i) (\underline{g}_i - g_i^\star) \}.
\end{aligned}
\end{equation}}
For scalars $a, b$, notice that $\max\{a, b\} = \frac{1}{2}\left({a + b + | a - b |}\right)$, and hence,
{\small \begin{equation}
\begin{aligned}
  2\LOC_i[\v{\pi}] 
  &= {(\pi_i - c_i) (\overline{g}_i + \underline{g}_i - 2 g_i^\star) + | \pi_i - c_i | (\overline{g}_i - \underline{g}_i)},
  \label{eq:revenue.loc}
\end{aligned}
\end{equation}}
where we make use of the fact that $\overline{g}_i \ge \underline{g}_i$. 
Notice that all terms depend on $\pi_i - c_i$. 
From \eqref{eq:rsced.kkt.stat.gr}, we infer
{\small 
\begin{equation}
    \pi_i - c_i = \overline{\gamma}^{\star}_i - \underline{\gamma}^{\star}_i + \sum_{k = 1}^K \left( \overline{\gamma}^{\star}_{k,i} - \underline{\gamma}^{\star}_{k,i} \right),
    \label{eq:revenue.diff}
\end{equation}}
that together with triangle inequality implies
{\small \begin{equation}
    | \pi_i - c_i | \le \overline{\gamma}^{\star}_i + \underline{\gamma}^{\star}_i + \sum_{k = 1}^K \left( \overline{\gamma}_{k,i} + \underline{\gamma}_{k,i} \right).
    \label{eq:revenue.absdiff}
\end{equation}}
Using \eqref{eq:revenue.diff} and \eqref{eq:revenue.absdiff} in \eqref{eq:revenue.loc} yields,
{\small \begin{equation*}
\begin{aligned}
   &2 \LOC_i[\v{\pi}] \le (\overline{\gamma}_i - \underline{\gamma}_i + \sum_{k = 1}^K \overline{\gamma}_{k,i} - \underline{\gamma}_{k,i})(\overline{g}_i + \underline{g}_i - 2 g_i^\star) \\
   &\qquad\qquad\quad + (\overline{\gamma}_i + \underline{\gamma}_i + \sum_{k = 1}^K \overline{\gamma}_{k,i} + \underline{\gamma}_{k,i})(\overline{g}_i - \underline{g}_i) \\
   &= 2 (\overline{\gamma}_i + \sum_{k = 1}^K \overline{\gamma}_{k,i}) (\overline{g}_i - g_i^\star) - 2 (\underline{\gamma}_i + \sum_{k = 1}^K \underline{\gamma}_{k,i})(\underline{g}_i - g_i^\star).
\end{aligned}
\end{equation*}}
Using \eqref{eq:rsced.kkt.cs.g}, we then have
{\small \begin{equation}\nonumber 
\begin{aligned}
  \LOC_i[\v{\pi}] &\le \left( \sum_{k = 1}^K \overline{\gamma}_{k,i} \right) (\overline{g}_i - g_i^\star) + \left( \sum_{k = 1}^K \underline{\gamma}_{k,i} \right)(g_i^\star - \underline{g}_i).
\end{aligned}
\end{equation}}
Taking the sum over all payments, we have
{\small \begin{equation}\nonumber 
\begin{aligned}
    \bone^\top \LOC[\v{\pi}] &\le \sum_{k = 1}^K \overline{\v{\gamma}}_k^\top (\overline{\v{g}} - \v{g}^\star) + \underline{\v{\gamma}}_k^\top (\v{g}^\star - \underline{\v{g}}),
\end{aligned}
\end{equation}}
where we have used positivity of each of the components to render the sum as a dot product. Then, applying the bound on \MS{} in \eqref{eq:ms.bound.3}, we have
{\small \begin{equation}\nonumber
\begin{aligned}
  \MS[\v{\pi}] - \bone^\top \LOC{}[\v{\pi}] \ge \sum_{k = 1}^K \overline{\v{\sigma}}_k \v{\Delta}_d + \overline{\nu}_k \voll^\top \v{\delta d}_k.
\end{aligned}
\end{equation}}
The result follows from the dual feasibility conditions and positivity of $\v{\Delta}_d, \voll, \v{\delta d}_k$.

%% file: KKT.tex
\begin{figure*}[!t]
\framebox[\linewidth][c]{
    \begin{minipage}{0.98\linewidth}
    \begin{itemize}
        \small
        \item Primal feasibility conditions: 
        \eqref{eq:RSCED.lp.nominal} --- 
        \eqref{eq::RSCED.lp.epigraph}.
        \item Dual feasibility conditions: $\v{\mu}, \underline{\v{\gamma}}, \overline{\v{\gamma}}, \v{\mu}_k^\DA{}, \v{\mu}_k^\SE{}, \underline{\v{\gamma}}_k, \overline{\v{\gamma}}_k, \underline{\v{\rho}}_k, \overline{\v{\rho}}_k, \underline{\v{\eta}}, \overline{\v{\eta}}, \underline{\v{\sigma}}_k, \overline{\v{\sigma}}_k, \underline{\nu}_k, \overline{\nu}_k \ge 0$ for all $k \in[K]$.
        \item 
        Stationarity conditions: For $k \in[K]$,
        \begin{subequations}
        \begin{gather}
          \v{c} - \lambda \bone + \v{H}^\top \v{\mu} - \underline{\v{\gamma}} + \overline{\v{\gamma}} + \sum_{k = 1}^K \left(\v{H}_k^\top (\v{\mu}_k^\DA{} + \v{\mu}_k^\SE{}) - \underline{\v{\gamma}}_k + \overline{\v{\gamma}}_k\right) = 0, \;
          \underline{\v{c}}_r - \underline{\v{\eta}} - \sum_{k = 1}^K \underline{\v{\rho}}_k  = 0, \;
          \overline{\v{c}}_r - \overline{\v{\eta}} - \sum_{k = 1}^K \overline{\v{\rho}}_k = 0, \label{eq:rsced.kkt.stat.gr} \\
          -\lambda_k \bone + \v{H}_k^\top \v{\mu}_k^\SE{} - \underline{\v{\gamma}}_k + \overline{\v{\gamma}}_k + \overline{\v{\rho}}_k - \underline{\v{\rho}}_k = 0, \;
          -\lambda_k \bone + \v{H}_k^\top \v{\mu}_k^\SE{} - \underline{\v{\sigma}}_k + \overline{\v{\sigma}}_k + \overline{\nu}_k \voll = 0, \label{eq:rsced.kkt.stat.d} \\
          1 - \sum_{k = 1}^K \overline{\nu}_k = 0, \;
          \frac{1}{1 - \alpha} p_k - \underline{\nu}_k - \overline{\nu}_k = 0. \label{eq:rsced.stat.cvar}
        \end{gather}
        \end{subequations}
        \item Complementary slackness conditions: 
        \begin{subequations}
        \begin{gather}
            \v{\mu}^\top [\v{H}(\v{g} - \v{d}) - \v{f}] = 0, \; \underline{\v{\gamma}}^\top [ \underline{\v{g}} - \v{g}] = 0, \;
            \overline{\v{\gamma}}^\top [\v{g} - \overline{\v{g}} ] = 0, \label{eq:rsced.kkt.cs.g} \\
            [\v{\mu}_k^\DA{}]^\top [\v{H}_k (\v{g} - \v{d}) - \v{f}_k^\DA{}] = 0, \;
            [\v{\mu}_k^\SE{}]^\top [\v{H}_k(\v{g} + \v{\delta g}_k - \v{d} + \v{\delta d}_k) - \v{f}_k^\SE{}] = 0, \label{eq:rsced.kkt.cs.flow} \\
            \underline{\v{\gamma}}_k^\top [ \underline{\v{g}} - \v{g} - \v{\delta g}_k] = 0, \;
            \overline{\v{\gamma}}_k^\top [ \v{g} + \v{\delta g}_k - \overline{\v{g}} ] = 0, \;
            \underline{\v{\rho}}_k^\top [ -\underline{\v{r}} - \v{\delta g}_k ] = 0, \;
            \overline{\v{\rho}}_k^\top [\v{\delta g}_k - \overline{\v{r}}] = 0, \label{eq:rsced.kkt.cs.dg} \\
            \underline{\v{\eta}}^\top \underline{\v{r}} = 0, \;
            \overline{\v{\eta}}^\top \overline{\v{r}} = 0, \; 
            \underline{\v{\sigma}}_k^\top \v{\delta d}_k = 0, \;
            \overline{\v{\sigma}}_k^\top [ \v{\delta d}_k - \v{\Delta}_d ] = 0, \; \underline{\nu}_k y_k = 0, \;
            \overline{\nu}_k [ y_k + z - \v{\voll}^\top \v{\delta d}_k ] = 0. \label{eq:rsced.kkt.cs.rdcvar}
        \end{gather}
        \end{subequations}
    \end{itemize}
    \end{minipage}
}
\caption{The KKT conditions for \eqref{eq:RSCED.lp}.}
\label{fig:rsced.kkt}
\end{figure*}